\pdfoutput=1
\RequirePackage{fix-cm}
\documentclass[natbib,smallextended]{svjour3}       
\smartqed  
\usepackage{graphicx}
\usepackage{url}
\usepackage{amsmath,amssymb,amsfonts}
\usepackage{wasysym}
\usepackage{color}
\usepackage{verbatim}
\usepackage[citecolor=blue,urlcolor=blue,breaklinks]{hyperref}
\newcommand{\be}{\begin{equation}}
\newcommand{\ee}{\end{equation}}
\newcommand{\mM}{\mathcal M}

\newcommand{\M}{M}

\newcommand{\Si}{\Sigma}
\newcommand{\Sr}{\Omega}
\newcommand{\Su}{\mathcal{S}}

\newcommand{\Rt}{\mathbb{R}^3}

\newcommand{\dv}{dv}

\newcommand{\dsf}{ds_0}
\newcommand{\ds}{ds}

\newcommand{\mf}{\mathcal{M}}

\newcommand{\mq}{M_{bh}}

\newcommand{\kif}{\bar\eta}
\usepackage{enumerate}

\journalname{Living Rev. Relativ.}
\begin{document}

\title{Geometrical inequalities bounding angular momentum and charges in General Relativity}

\titlerunning{Geometrical inequalities bounding angular momentum and charges in GR}        

\author{Sergio Dain \and Mar\'ia Eugenia Gabach-Clement\\
}

\institute{Facultad de Matem\'atica, Astronom\'ia y F\'isica, 
Universidad Nacional de C\'ordoba, 
Instituto de F\'isica Enrique Gaviola, Consejo Nacional de Investigaciones Cient\'ificas y Tecnol\'ogicas
Argentina.
              \email{gabach@famaf.unc.edu.ar}    
           }

\date{Received: date / Accepted: date}

\maketitle

\begin{abstract}
Geometrical inequalities show how certain parameters of a physical system set restrictions on other parameters. For instance, a black hole 
of given mass can not rotate too fast, or an ordinary object of given size can not have too much electric charge. In this article we 
are interested in bounds on the angular momentum and electromagnetic charges, in terms of total mass and size. We are mainly concerned
with inequalities for black holes and ordinary objects. The former  are the most studied systems in this context in General Relativity, 
and where most results have been
found. Ordinary objects, on the 
other hand, present numerous challenges and many basic questions concerning geometrical estimates for them are still unanswered. 

We present the many results in 
these areas. 
We make emphasis in  identifying the mathematical conditions that lead to such estimates, both for black 
holes and ordinary objects.
\keywords{Geometrical inequalities \and Black holes \and Ordinary objects}
\end{abstract}

\newpage
\begin{flushright}\textit{The important moments in life are the ones we share with others.}
 
\textit{ Sergio Dain}
\end{flushright}

\vspace{3cm} 
\begin{center}
This article is dedicated to the memory of\\ 
Sergio Dain and Marcus Ansorg
\end{center}
\newpage

\setcounter{tocdepth}{3}
\tableofcontents

\newpage

\section{Introduction}\label{sec:intro}
Geometrical inequalities in General Relativity have always played a key role in understanding some physical systems. The basic questions behind these inequalities are the following. 
What are the reasons that such inequalities do exist at all? Another, more humble 
but more practical and hopefully illuminating question is about the elements in Einstein theory, that produce such inequalities.
In other words: Why should we expect such inequalities, and where do they come from? The straightforward answer is that there are a number 
of different objects predicted by the theory that live in different regimes. 

This is clear when one thinks about one of the most important solutions to Einstein equations: Black holes.
In black hole systems, there are two natural limits to consider. One of them is 
the maximum mass a star can have, beyond which it collapses into a black hole. This problem was 
addressed in the 1930s  by Chandrasekhar (see \citealp{Chandrasekhar:1985kt} and the references therein). The other is the
maximum charge and/or  angular momentum a black hole can have, beyond which it becomes a naked singularity. This problem arose
after \cite{Reissner16} and \cite{Nordstrom18} found the solution  describing a static, spherically symmetric, electrically 
charged object.

These thresholds in the physical parameters values give rise to geometrical inequalities. In more general terms, we can naively imagine a 
function $f$ depending 
on the physical parameters of the system, like the mass $M$, size $ R$, angular momentum $J$, electromagnetic charge $Q$, etc., 
denoted by 
$f:=f(M,R,J,Q, ... )$, such that when $f\in [f_{obj}^-,f_{obj}^+]$, the system describes an 
non black hole ordinary object, when $f\in [f_{bh}^-,f_{bh}^+]$, it describes a black hole, and when $f\in [f_{nak}^-,f_{nak}^+]$, a naked singularity.
To illustrate this point, consider a rotating object of surface area $A$, angular momentum $J$ and characteristic radii $ R_1, R_2$. Then  if we take
the function $f$ as the angular momentum $J$, the parameter space can be divided in the following form
\begin{eqnarray}\label{regimobj}
\mbox{Ordinary object} \quad& \Rightarrow& \quad0\leq|J|\leq R_1^2   \\\label{regimbh}
\mbox{Black hole}\quad& \Rightarrow& \quad R_2^2\leq |J|\leq A/8\pi \\\label{regimnak}
\mbox{Naked singularity}\quad& \Rightarrow& \quad A/8\pi<|J|. 
\end{eqnarray}
We emphasize that this is a very rough and  overly simplified picture of what the geometrical inequalities found so far 
actually say, and of what to expect for more general systems. But, as we see in Sects.~\ref{sec:AJ} and \ref{sec:ordinary}, the division of the parameter space showing the different regimes in which the system 
can exist, is what one actually finds in some cases.

In this article we address the first two regimes, given in \eqref{regimobj}  and \eqref{regimbh}. The latter, and the frontier 
between black holes and naked singularities, is the original and main motivation for the geometrical inequalities
presented here. Therefore, we explore it in what follows with a paradigmatic black hole solution.

Consider the Kerr--Newman black hole with mass $M$, angular momentum $J$ and
electric charge $Q$ (see \citealp{Wald84}). The area $A$ of the horizon is given by 
\begin{equation}
  \label{eq:19}
  A=4\pi \left(2M^2-Q^2+2M \sqrt{d}  \right), \quad d= M^2-Q^2-\frac{J^2}{M^2}.
\end{equation}
The equality (\ref{eq:19})  implies the following three important inequalities
among the parameters:
\be\label{eq:penrose1}
\sqrt{\frac{A}{16\pi}}\leq M,
\ee
\be  \label{eq:24i}
 \frac{Q^2+\sqrt{Q^4+4J^2}}{2} \leq M^2,
 \ee
 \be
  \label{eq:13}
  4\pi \sqrt{Q^4+4J^2} \leq A.
  \ee
  These inequalities saturate in the two relevant limit values for the
parameters: the Schwarzschild black hole given by $Q=J=0$ (where
(\ref{eq:penrose1}) reaches the equality) and the extreme Kerr--Newman black hole
given by $d=0$ (where the inequalities (\ref{eq:24i}) and (\ref{eq:13}) reach
the equality). Note that inequality (\ref{eq:24i}) is equivalent, by a simple
computation, to the condition $d\geq 0$.  

It is important to recall that the
Kerr--Newman metric is a solution of Einstein electrovacuum equations for any
choice of the parameters $(M,J,Q)$.  However, it represents a black hole (and
hence the area $A$ of the horizon is well defined) if and only the inequality
(\ref{eq:24i}) holds. Otherwise the spacetime contains a naked singularity.

Above, we have derived inequalities (\ref{eq:penrose1})--(\ref{eq:13}) from a very
particular exact solution of Einstein equations: the Kerr--Newman stationary
black hole. However, remarkably, these inequalities remain valid (under
appropriate assumptions) for fully dynamical black holes. Moreover, they are
deeply connected with properties of the global evolution of Einstein equations,
in particular with the cosmic censorship conjecture.  

The inequalities  (\ref{eq:penrose1})--(\ref{eq:13})  can be divided
into two groups:

\begin{enumerate}
\item $\sqrt{\frac{A}{16\pi}}\leq M$: the area appears as lower bound.

\item $\frac{Q^2+\sqrt{Q^4+4J^2}}{2} \leq M^2$ and $4\pi \sqrt{Q^4+4J^2} \leq
  A$ : the angular momentum and the charge appear as lower bounds.
\end{enumerate}
This division seems rather unnatural at first, due to the the quantities involved being the same and the fact that inequalities
\eqref{eq:penrose1} (in the 
first group) and \eqref{eq:13} (in the second group) look like intermediate inequalities of  \eqref{eq:24i} (in the second group).
However, at the moment the division makes sense because the mathematical methods used to
study these two groups are in general different. We expect that in the future new
connections will appear between all these inequalities.

As the title of this article suggests, we will focus on the second group. For dynamical black
holes, the inequality (\ref{eq:penrose1}) in the first group is the Penrose
inequality. There exists already an excellent and up to date review on this subject by \cite{Mars:2009cj}.

Furthermore, the inequalities can be also classified as global or quasilocal.  We  explain this distinction in more detail in section
\ref{sec:physical-quantities}. Roughly speaking, the total mass $M$ is a global
quantity (i.e. it depends on the whole spacetime), in contrast the area $A$, the
charge $Q$ and the angular momentum $J$ are quasilocal quantities, they depend
on bounded regions of the spacetime.  That is, (\ref{eq:penrose1}) and
(\ref{eq:24i}) are global inequalities, and (\ref{eq:13}) is a purely
quasilocal inequality.  Global inequalities can be interpreted as refinements of
the positive mass theorem in the presence of a black hole.

Our main interest is the study of inequalities 
(\ref{eq:penrose1})--(\ref{eq:13}) for
dynamical black holes and also in related or more general situations like
stationary black holes with surrounding matter fields, ordinary objects, higher
dimensions and alternative theories of gravity. However, in this article
there are some topics that are left uncovered. Some of them are: 

\begin{itemize}
\item \textit{Geometrical inequalities involving quasilocal mass}.
This is a very broad subject as there are many different notions of quasilocal mass and energy. The problem of determining a unique 
appropriate notion that will give general useful and representative geometrical inequalities is open. There is a beautiful review by 
\cite{Szabados04}, on quasilocal quantities, and in particular, quasilocal mass, that discusses this issue 

\item \textit{Geometrical inequalities for black holes in higher dimensions and within alternative theories of gravity}. This topic has been growing 
during the last years and many very interesting results have been
obtained. See the work of \cite{Gibbons06, Gibbons:1998zr}, \cite{Hollands:2011sy}, \cite{Yazadjiev:2012bx, Yazadjiev:2013hk}, \cite{Fajman:2013ffa}, \cite{Alaee:2017ygv, Alaee:2016jlp, Alaee:2015pwa, Alaee:2015gga, Alaee:2014tfa}, \cite{Rogatko:2017oea, Rogatko:2014yoa}, and references therein.

\end{itemize}
The results presented in this article can be grouped essentially in three parts. The first two parts concern black holes. Global
inequalities of the form (\ref{eq:24i}) are reviewed in Sect.~\ref{sec:global}, and quasilocal inequalities of the form
(\ref{eq:13}) are presented in Sect.~\ref{sec:QL}. In Sect.~\ref{relation} we discuss a partial relation between the global and the  
quasilocal problems. The third part, in Sect.~\ref{sec:objects} addresses geometrical 
inequalities for non black hole objects. 
The mathematical methods used to study the various problems are
similar in many ways but the physical
implications and scopes of these types of inequalities appear to be very
different, we will address this issue  in the following sections. 

There are also a number of articles reviewing the subject of geometrical inequalities that include some of the results presented here. 
They were written by \cite{Dain:2011mv, dain12, Dain:2014xda}, and \cite{Jaram12} with slightly different approaches and focuses.

\subsection{Motivation from stationary black holes}
\label{sec:phys-interpr-stat}

Before discussing the general setting, it is important to analyze, in a
heuristic way, the physical meaning of the inequalities
(\ref{eq:penrose1})--(\ref{eq:13}) for stationary black holes.

Let us begin with inequality (\ref{eq:penrose1}). This inequality describes the
most basic property of a black hole, that is, its mass is concentrated
in a small region of space. More precisely, in terms of the areal radius $ R:=\sqrt{A/4\pi}$, 
this inequality is expressed as
\begin{equation}
  \label{eq:25}
  \frac{ R}{2}\leq M.
\end{equation}
Inequality (\ref{eq:25})  can be interpreted as a weak version of the
Hoop conjecture \citep{thorne72}, in which the area is taken as a measure of the size of the black
hole. See \cite{Senovilla:2007dw} for alternative formulations and references 
to previous works on the conjecture, and the more recent articles by \cite{Malec:2015oza}, 
\cite{Yoshino:2007yb}, \cite{Gibbons:2009xm}, \cite{Khuri:2009dt}, \cite{Murchadha:2009lav}, \cite{Hod:2015iig}, and \cite{Cvetic:2011vt}.

Consider the second inequality, (\ref{eq:24i}). Using a mixture of classical and
relativistic equations, in the following we will argue that this inequality is
essentially a consequence of (\ref{eq:25}). Take a sphere of
radius $R$ with constant electric density and total charge $Q$. The classical
electromagnetic energy of this sphere is given by
\begin{equation}
  \label{eq:18}
  W_Q =\frac{3}{5}\frac{Q^2}{R}. 
\end{equation}
In addition, suppose that the sphere has mass $M$, constant density and it
rotates with constant angular velocity. The Newtonian kinetic energy of the
sphere is given by
\begin{equation}
  \label{eq:20}
  W_J = \frac{5}{4}\frac{J^2}{MR^2},
\end{equation}
where  $J$ is the angular momentum of the sphere.  

Assume that the sphere collapses and forms a black hole. The total  mass $M$ of the black
hole should be greater than the sum of the energies
\begin{equation}
  \label{eq:22}
 W_Q+W_J \leq M.
\end{equation}

We use the inequality (\ref{eq:25}) to bound the radius $R$ by the mass in the
energies $W_Q$ and $W_J$, we obtain
\begin{equation}
  \label{eq:27}
 \frac{3}{10}\frac{Q^2}{M}  + \frac{5}{16}\frac{J^2}{M^3} \leq
 \frac{3}{5}\frac{Q^2}{R}  + \frac{5}{4}\frac{J^2}{MR^2}=  W_Q+W_J.
\end{equation}
Hence using (\ref{eq:22}) we finally get
\begin{equation}
  \label{eq:24}
  \frac{Q^2}{M}+\frac{J^2}{M^3}   \lesssim  M.
\end{equation}
Note that inequality (\ref{eq:24}) is equivalent to the condition $d\geq 0$
given in (\ref{eq:19}) and hence we recover the inequality (\ref{eq:24i}).  The
symbol $\lesssim$ 
in (\ref{eq:24}) means that in the left hand side of this
equation we have approximated all the numerical factors that appear in equation
(\ref{eq:27}) by one. These numbers  depend  on the specific matter
model we have chosen for the sphere: i.e. constant charge and mass density. We
can not expect, by this kind of argument, to obtain the precise numerical
factors involved in the inequality (\ref{eq:24i}), only the order of
magnitude. However, remarkably, we have obtained the correct functional
dependence on the parameters.

Finally, consider the last inequality (\ref{eq:13}).  In the following, using
thermodynamics arguments, we will argue that this inequality is a consequence
of the inequality (\ref{eq:24i}) and the existence of extreme black holes
(i.e. black holes with non-zero area that saturates (\ref{eq:24i})).

Consider a general stationary black hole which is not necessarily Kerr--Newman,
for example a stationary black hole surrounded by a ring of matter.  Assume that
there exists a function of the form $A(M,J,Q)$ that relates the parameters of
the black hole. We can include more parameters to this equation without
altering the following argument.  If we identify the area $A$ as the entropy of
the black hole, then, in the thermodynamical language, the function
$A(M,J,Q)$ would be identified as the fundamental equation of the system. Its existence is
one of the postulates of Thermodynamics (see, for example, \citealp{callen85}).
The inverse of the temperature of the system is defined as the partial
derivative of $A(M,J,Q)$ with respect to $M$, and it is a positive quantity.  
That is, if we define
$\kappa$ by
\begin{equation}
  \label{eq:6v}
  \frac{\partial A}{\partial M}  = \frac{8 \pi}{\kappa},
\end{equation}
we have $\kappa \geq 0$.  Hence, $A(M,J,Q)$ is an increasing function of $M$
for fixed $J$ and $Q$.

Assume, in addition, that we have a lower bound for the mass $M$ in terms of
$J$ and $Q$ like the inequality (\ref{eq:24i}). We do not need the particular
form given by (\ref{eq:24i}), we assume some general inequality of the form
\begin{equation}
  \label{eq:M0}
  M\geq M_0(J,Q),
\end{equation}
where $M_0(J,Q)$ is a strictly positive given function.  Consider the function
$A(M,J,Q)$ for fixed $J$ and $Q$. The bound (\ref{eq:M0}) implies that for $M$
only values with $M\geq M_0$ are allowed. Since $A$ is an increasing function
of $M$ we obtain
\begin{equation}
  \label{eq:7b}
  A(M,J,Q)\geq A(M_0,J,Q)=A_0(J,Q),
\end{equation}
where we have defined the function $A_0(J,Q)$  by
\begin{equation} 
  \label{eq:23}
  A_0(J,Q)=A(M_0(J,Q),J,Q).
\end{equation}
In order to obtain from (\ref{eq:7b}) a non-trivial inequality we need to assume that
$A_0>0$ (in principle we could have $A_0$ identically zero). That is, we need
to assume the existence of non trivial extreme black holes:  black holes
for which the bound (\ref{eq:M0}) is saturated and have non-zero area. 
We have obtained the inequality (\ref{eq:7b}), which has the same form as
(\ref{eq:13}). Of course, in the Kerr--Newman case we have
\begin{equation}
  \label{eq:40}
  M_0(J,Q)=\frac{Q^2+\sqrt{Q^4+4J^2}}{2},  \quad A_0(J,Q)=  4\pi \sqrt{Q^4+4J^2}, 
\end{equation}
and hence (\ref{eq:7b}) has exactly the same form as (\ref{eq:13}).

Given that the function $A(M,J,Q)$ ends at the value $A_0$, one can ask what
happens at that point. For the extreme black hole the temperature is zero,
hence the derivative (\ref{eq:6v}) is infinite and then the function $A(M,J,Q)$
can not be extended in any smooth way beyond the point $A_0$. This can be
explicitly checked in the Kerr--Newman case given by (\ref{eq:19}).

In summary, these informal results show that inequalities \eqref{eq:penrose1}-\eqref{eq:13} are motivated 
by the Kerr family of stationary black holes. But there is another interesting observation in the above arguments. Namely, that Penrose inequality \eqref{eq:penrose1} 
implies the global 
inequality \eqref{eq:24i} and the global inequality implies the quasilocal inequality \eqref{eq:13}
\be\label{implications}
M\geq\sqrt{\frac{A}{16\pi}}\Rightarrow\quad M^2\geq \frac{Q^2+\sqrt{Q^4+4J^2}}{2} \Rightarrow\quad A\geq  4\pi \sqrt{Q^4+4J^2} .
  \ee
We note that we use the 
\textit{uncharged} and \textit{non-rotating} version of Penrose inequality (see \citealp{Mars:2009cj} for a review on the subject and 
\citealp{Khuri:2014wqa} for a recent result on the charged Penrose inequality for multiple black holes).  The first implication in 
\eqref{implications} may be relevant when extending the Penrose inequality to include angular momentum and charge. As we will see, the
treatments of \eqref{eq:24i} and \eqref{eq:13} are similar in some ways and in fact, a version of the second implication is 
obtained in the general dynamical scenario (see Sect.~\ref{relation}).

The implications \eqref{implications} also put the Penrose inequality in a very especial place as being, in a sense, more basic than the 
other inequalities. We come back to this issue in Sect.~\ref{sec:phys-heur-argum} where it is deduced from standard arguments in 
collapse scenarios, and in Sect.~\ref{sec:newt} where it is a result of Newtonian considerations
with the only condition that
the velocity of any particle should be smaller than or equal to the velocity of light.

\subsection{Heuristic arguments in dynamical black hole regimes}
\label{sec:phys-heur-argum}

The extension of the inequality \eqref{eq:penrose1} for dynamical black holes
was done by \cite{Penrose73} using a remarkable physical argument
that connects global properties of the gravitational collapse with geometric
inequalities on the initial conditions.  We briefly review this argument below
(see also \citealp{Mars:2009cj}, \citealp{dain12}, \citealp{Dain:2014xda} and references
therein)

We will assume that the following statements hold in a gravitational collapse:
\begin{itemize}

\item[(i)]  Gravitational collapse results in a black hole (weak cosmic
censorship). 

\item[(ii)]  The spacetime settles down to a stationary final
state. Furthermore, at some finite time after the collapse,  all the non
electromagnetic  matter fields  have
fallen into the black hole and hence the exterior region is
electro-vacuum. 
\end{itemize}
Conjectures (i) and (ii) constitute the standard picture of the gravitational
collapse. 

The black hole uniqueness theorem implies that the final state postulated in
(ii) is given by the Kerr--Newman black hole (we emphasize however that many
important aspects of  black hole uniqueness still remain open, see
\cite{Chrusciel:2008js} for a recent review on this problem).  Let us denote by
$M_0$, $A_0$, $J_0$ and $Q_0$ the mass, area, angular momentum and charge of
the remainder Kerr--Newman black hole.  These quantities will, of course,
satisfy the three inequalities (\ref{eq:penrose1})--(\ref{eq:13}).

Let us consider an initial data set  for a
gravitational collapse such that the collapse has already occurred on the data.
That means that the initial spacelike surface $\Si$ intersects the event horizon of the
black hole. The intersection is a spacelike, closed, 2-surface denoted by $\Su$
with area $A(\Su)$.  Let $M$ be the total mass of the spacetime
defined by (\ref{eq:massadm}).

From the black hole area theorem  \citep{Hawking71, Chrusciel:2000cu} we have that the area of the black hole increases
with time and hence
\begin{equation}
  \label{eq:15}
  A \leq A_0.
\end{equation}
Since gravitational waves carry positive energy, the total mass of the
spacetime should be bigger than the final mass of the black hole
\begin{equation}
  \label{eq:4v}
  M_0 \leq  M.
\end{equation}
The difference $M-M_0$ is the total amount of gravitational radiation emitted
by the system. Combining (\ref{eq:15}) and (\ref{eq:4v}) and the fact that the
remainder black hole satisfies the inequality  (\ref{eq:penrose1}), namely
\begin{equation}
  \label{eq:7v}
  \sqrt{\frac{A_0}{16\pi}} \leq M_0,
\end{equation}
we finally conclude that
\begin{equation}
  \label{eq:23v}
   \sqrt{\frac{A}{16\pi}}\leq M.
\end{equation}
There is still an important issue to be discussed: how to estimate the area $A(\Su)$ in
terms of geometrical quantities that can be  computed from the initial
conditions. Recall that in order to know the location of the event horizon the
entire spacetime is needed.  Assume that the surface $\Si$ contains a future
trapped 2-surface $\Su_0$. By a
general result on black hole spacetimes \citep{Penrose65, Hawking71}, we know that the surface $\Su_0$
should be contained in $\Su$. But that does not necessarily mean that the
area of $\Su_0$ is smaller than the area of $\Su$.  Consider all surfaces
$\tilde \Su$ enclosing $\Su_0$. Denote by $A_{\min}(\Su_0)$ the
infimum of the areas of all such surfaces. Then we clearly have that
$A(\Su)\geq A_{\min}(\Su_0)$.  The advantage of this construction is that
$A_{\min}(\Su_0)$ is a quantity that can be computed from  the Cauchy surface
$\Si$. Using this inequality 
we finally obtain the Penrose inequality
\begin{equation}
  \label{eq:58}
  \sqrt{\frac{ A_{\min}(\Su_0)}{16\pi}} \leq M.
\end{equation}
In the time symmetric case (i.e. when $K_{ij}=0$) an important simplification
occurs. For that case,  a marginally trapped outermost surface  is a minimal surface (see Sect.~\ref{sec:BH} for definitions), and hence
we do not need to consider the family of enclosing surfaces. For further
discussion we refer to \cite{Mars:2009cj} and references therein.

The key point in the previous argument is that there exist simple inequalities that relate the quantities $M$ and $A$ on the 
initial conditions with the quantities $M_0$ and $A_0$ on the remainder black holes, where the geometrical inequalities are satisfied. 
For the second inequality (\ref{eq:24i}) we need to consider the electric charge and the angular momentum.  The problem is that there is no  simple inequality, like \eqref{eq:4v}, 
that relates the total electric charge and angular momentum of the initial data
$(Q_\infty,J_\infty)$ with the corresponding quantities of the final, Kerr--Newman black hole $(Q_0,J_0)$. We need additional assumptions. 

Suppose that in the exterior of the black hole the matter fields are not charged.  That is
\begin{equation}
  \label{eq:29}
   \nabla_\mu T^{EM\, \mu\nu}=0,
\end{equation}
on $\Si\setminus \Su$, where $T^{EM\, \mu\nu}$ is the energy momentum tensor field of electromagnetism. Then, the electric charge is conserved 
in the exterior region (see Sect.~\ref{sec:physical-quantities}) and hence we have
\begin{equation}
  \label{eq:21b}
  Q_\infty=Q_0
\end{equation}
Using (\ref{eq:21b}), the explicit expression for the Maxwell energy momentum tensor (\ref{eq:4}), and the fact that the remainder black hole satisfies
\begin{equation}
  \label{eq:31}
  |Q_0|\leq M_0,
\end{equation}
we get
\begin{equation}
  \label{eq:30}
  |Q_\infty|\leq M.
\end{equation}
That is, we have obtained the dynamical version of the inequality (\ref{eq:24i}) for the case $J=0$. 
Note that we have not used the area theorem. We  see in section
\ref{sec:global} that the assumption that the matter fields are non-charged in
the black hole exterior region can be slightly relaxed: it can be assumed that
the electric charge density is small with respect to the mass density.

To relate the initial angular momentum $J_\infty$ with the final angular
momentum $J_0$ is much more complicated. Angular momentum is in general
non-conserved. There exists no general simple relation between the total angular
momentum $J_\infty$ of the initial conditions and the angular momentum $J_0$ of the
final black hole. For example, a system can have $J_\infty=0$ initially, but collapse
to a black hole with final angular momentum $J_0\neq 0$. We can imagine that on
the initial conditions there are two parts with opposite angular momentum, one
of them falls into the black hole and the other escapes to infinity. Axially
symmetric vacuum spacetimes constitute a remarkable exception because the
angular momentum is conserved in electrovacuum. That is, we have
\begin{equation}
  \label{eq:59b}
  J_\infty=J_0.
\end{equation}
For a discussion of this conservation law in detail see Sect.~\ref{sec:physical-quantities}. Using (\ref{eq:59b}), (\ref{eq:21b}), 
\eqref{eq:4v} and \eqref{eq:40} we
finally obtain the dynamical version of the inequality (\ref{eq:penrose1})
\begin{equation}
  \label{eq:32}
   \frac{Q_\infty^2+\sqrt{Q_\infty^4+4J_\infty^2}}{2} \leq M^2.
\end{equation}
We emphasize that the  inequality (\ref{eq:32}) holds under the assumption of
axial symmetry and electrovacuum in the exterior region of the black hole.

We have seen that the two inequalities (\ref{eq:penrose1}) and (\ref{eq:24i})
extend to the dynamical regime in the forms (\ref{eq:58}) and (\ref{eq:32}).
These inequalities are global because the mass $M$ is the total mass of the
spacetime. Whether the quantities in these inequalities can be replaced by an
appropriate defined black hole quasilocal mass and angular momentum (without any
symmetry assumption) is unknown.

Penrose argument is remarkable because starting from conjectures (i) and (ii) one is able
to deduce inequalities that can be written purely in terms of the initial
conditions. That is, the inequalities do not involve the unknown parameters
$M_0$, $Q_0$, $J_0$, $A_0$ of the remainder black hole. A counter example to
inequalities (\ref{eq:58}) or (\ref{eq:32}) will be a counter example of the
conjectures (i) or (ii). On the other hand, the proof of such inequalities
gives indirect evidences of the validity of the conjectures (i) and (ii).  In
that sense, the physical heuristic argument is quite strong: in either
direction (i.e. if the inequalities are valid or not) it provides highly
non-trivial new insight. In contrast, the physical heuristic arguments for the
validity of the quasilocal inequality (\ref{eq:13}) in the dynamical regime are
less conclusive.

The argument we present in Sect.~\ref{sec:phys-interpr-stat} uses
thermodynamics, and hence its validity outside equilibrium is not
clear. Nevertheless, the quantities involved in (\ref{eq:13}) are, as we have
seen in Sect.~\ref{sec:physical-quantities}, well defined quasilocal
quantities in the full dynamical regime (in the case of angular momentum we
need the additional assumption of axial symmetry).  Consider a
stationary black hole that satisfies inequality (\ref{eq:13}) (we can
assume instead that it satisfies the more general version (\ref{eq:7b}), the
following argument will be identical). We make a perturbation to this
stationary black hole that preserves the charge and angular momentum of the
black hole. For example, a vacuum axially symmetric perturbation will have this
property.  Physically, the stationary black hole will absorb axially symmetric
gravitational waves without changing its charge and angular momentum. The area,
however, will increase. That means that the same inequality (\ref{eq:13})
will be satisfied for this dynamical black hole in the future. Slightly more
general, if we have a system of multiple black holes such that in the past they
can be approximated by isolated stationary black holes and such that the whole
spacetime is axially symmetric and electrovacuum, then, by the same argument we
expect that the inequality (\ref{eq:13}) will be satisfied for each individual
black hole. Head on collision of black holes is an example of such situation.
Hence, by this simple argument, we expect a large class of dynamical black
holes for which the inequality (\ref{eq:13}) is satisfied. However, it is not
obvious how to rule out black holes that can not be treated as continuous
deformation of stationary black holes (although perhaps such situation does not
occur). For that cases, we can argue as follows. If the inequality is not
satisfied for a dynamical black hole, then it should be possible to perturb it
in the same way as above, increasing the area and preserving the angular
momentum and charge. There is in principle no physical restriction to 
how much the area changes, as long as it increases. Hence, it should be
possible to increase the area until the equality in (\ref{eq:13}) is reached. The
arguments presented in Sect.~\ref{sec:phys-interpr-stat} suggest that the
equality is reached only for extreme black holes. And there are well known
physical arguments which suggest that extremal black holes can not be produced
in a finite process \citep{Bardeen:1973gs, Carroll:2009maa}.

Finally, we review  the original argument in favor of   (\ref{eq:13})
presented in  \cite{dain10d} (see also \citealp{dain12}). 
Consider the formula (\ref{eq:19}) for the horizon area of the Kerr--Newman
black holes. From this expression, we can write the mass in terms of the other
parameters. Since in axial symmetry we have a well defined quasilocal
definition of angular momentum (Sect.~\ref{sec:physical-quantities}) we can formally define the quasilocal mass of a
black hole by the same expression as the mass for the Kerr--Newman black hole
but replacing the parameters by its quasilocal definition, namely
\begin{equation}
  \label{eq:32b}
  \mq(\Su)= \sqrt{\frac{A(\Su)}{16\pi}+ \frac{Q(\Su)^2}{2}+\frac{\pi(Q(\Su)^4 +4J_T(\Su)^2)}{A(\Su)}}.
\end{equation}
Note that in (\ref{eq:32b}) we have used the total quasilocal angular momentum
(i.e. gravitational plus electromagnetic) 
(see definition (\ref{eq:43}) in Sect.~\ref{sec:quasilocalq}). The relevant question is: does $\mq$ describe the quasilocal mass of
a non-stationary axially symmetric black hole? This question is closely
related to the validity of the inequality   (\ref{eq:13}) in the dynamical regime. In order to
answer it, let us analyze the evolution of $\mq$.

By the area theorem, we know that the horizon area will increase. If we assume
axial symmetry and electrovacuum, then the total angular momentum and the
charge will be conserved at the quasilocal level as we see in section
\ref{sec:physical-quantities}. On physical grounds, one would expect that in
this situation the quasilocal mass of the black hole increases with the
area, since there is no mechanism at the classical level to extract mass from
the black hole.  In effect, the only way to extract mass from a black hole is
by extracting angular momentum through a Penrose process \citep{Penrose:1971uk, Christodoulou70}.  But angular momentum
transfer is forbidden in electrovacuum axial symmetry.  Then, one would expect
both the area $A$ and the quasilocal mass $\mq$ to monotonically
increase with time.

Let us take a time derivative of $\mq$ (denoted by a dot).  To analyze this, it
is illustrative to write down the complete differential, namely the first law of
thermodynamics \citep{Bardeen:1973gs}
\begin{equation}
  \label{eq:55}
  \delta \mq=  \frac{\kappa}{8 \pi} \delta A + \Omega_H \delta J_T + \Phi_H
  \delta Q,
\end{equation}
where
\begin{equation}
  \label{eq:56}
  \kappa= \frac{1}{4 \mq} \left(1- \left(\frac{4\pi}{A}\right)^2 (Q^4+4J_T^2)\right),  
\end{equation}
\be\label{eq:56a}
 \Omega_H= \frac{4\pi J_T}{A\mq}, \quad  \Phi_H= \frac{4\pi (\mq+\sqrt{d}) Q}{A},
\ee
where $\mq$ is given by (\ref{eq:32}) and $d$ (defined in equation
 (\ref{eq:19}) ) is written in terms of $A$ and $J_T$ and $Q$ as
\begin{equation}
  \label{eq:17}
  d= \frac{1}{\mq^2}\left(\frac{A}{16\pi}\right)^2 \left(1-(Q^4+4J_T^2)\left(\frac{4\pi}{A} \right)^2 \right)^2.
\end{equation}

Under our assumptions, from the formula
\eqref{eq:32b} we obtain
\begin{equation}
  \label{eq:34}
  \dot{M}_{bh}  = \frac{\kappa }{8\pi}\dot A , 
\end{equation}
were we have used that the total angular momentum $J_T$ and the charge $Q$ are
conserved. Since, by the area theorem, we have
\begin{equation}
  \label{eq:9}
  \dot A \geq 0,
\end{equation}
the time derivative of $\mq$ will be positive (and hence the mass $\mq$ will
increase with the area) if and only if $\kappa \geq 0$, that is 
\begin{equation}
  \label{eq:5}
  4\pi \sqrt{Q^4+4J_T^2} \leq  A.
\end{equation}
Then, it is natural to conjecture that \eqref{eq:5} should be satisfied for any
black hole in axially symmetry.  If the horizon violates
\eqref{eq:5} then in the evolution the area would increase but the mass $\mq$
would decrease. This would indicate that the quantity $\mq$ has not the desired
physical meaning.  Also, a rigidity statement is expected. Namely, the equality
in \eqref{eq:5} is reached only by the extreme Kerr black hole given by the
formula
\begin{equation}
  \label{eq:6}
  A = 4\pi\left (\sqrt{Q^4+4J_T^2 }\right). 
\end{equation}
The final picture is that the size of the black hole is bounded from below by
the charge and angular momentum, and the minimal size is realized by the
extreme Kerr--Newman black hole.  This inequality provides a remarkable
quasilocal measure of how far a dynamical black hole is from the extreme case,
namely an `extremality criteria' in the spirit of \cite{Booth:2007wu}, although
restricted only to axial symmetry.   Note also that the inequality
\eqref{eq:5} allows to define, at least formally, the positive surface gravity
density (or temperature) of a dynamical black hole by the formula (\ref{eq:56})
(see \citealp{Ashtekar03, Ashtekar02} for a related discussion of the
first law in dynamical horizons).

If inequality \eqref{eq:5} is true, then we have a non trivial monotonic
quantity (in addition to the black hole area) $\mq$ in electro-vacuum
\begin{equation}
  \label{eq:10b}
 \dot{M}_{bh} \geq 0.
\end{equation}

It is important to emphasize that the physical arguments presented above in
support of \eqref{eq:5} are certainly weaker in comparison with the ones behind
the Penrose inequalities (\ref{eq:58}), (\ref{eq:22}) and (\ref{eq:23}).  A
counter example of any of these inequalities would prove that the standard picture
of the gravitational collapse is wrong. On the other hand, a counter example of
\eqref{eq:5} would only prove that the quasilocal mass \eqref{eq:32} is not
appropriate to describe the evolution of a non-stationary black hole.  One can
imagine other expressions for quasilocal mass, maybe more involved, in axial
symmetry.  On the contrary, reversing the argument, a proof of \eqref{eq:5}
will certainly suggest that the mass \eqref{eq:32} has physical meaning for
non-stationary black holes as a natural quasilocal mass (at least in axial
symmetry). Also, the inequality \eqref{eq:5} provides a non trivial control of
the size of a black hole valid at any time.

\subsection{Motivation from Newtonian objects}\label{sec:newt}
Geometrical inequalities for ordinary matter fields have gained much interest in the recent years. These inequalities are not expected to 
arise in Newtonian theory, unless some specific systems or matter fields are considered, which have intrinsic restrictions. With 
the state of  the subject at  present, we can not say that geometrical inequalities for objects are produced solely by Einstein equations. Other 
ingredients must be considered. In particular, one needs an analog of the variational characterization of extreme black holes, that we showed in the 
previous sections.

What is interesting is that one obtains geometrical inequalities from Newtonian considerations if they are supplemented with a key 
ingredient from General Relativity, namely, that any velocity in the system is not greater than the velocity of light. 
We discuss these inspirational inequalities in what follows.

We first consider the most basic argument in favor of Hoop inequality when we look at an ordinary material object in Newtonian theory. 
For simplicity, take it to be a spherical, static object of (quasilocal) mass $m$ and radius $R$. Then the escape velocity from the 
surface of this object is
\be\label{escape}
v_{esc}=\sqrt{\frac{2m}{R}}.
\ee
Now, assuming the velocity of any particle escaping from the object is not greater than the velocity of light, $v_{esc}\leq 1$, we obtain
\be\label{escape2}
m\leq \frac{R}{2}
\ee
for the object, which agrees with \eqref{eq:25}. Interestingly, there is a related conjectured inequality introduced by \cite{Yodzis:1973gha} and known 
as the trapped surface conjecture. It states that is the mass $m$ enclosed in a region of size $R$ does not satisfy
\be\label{escape3}
m\leq R
\ee
 then the region must be trapped.

Now we seek a quasilocal relation between angular momentum and size. We follow \cite{Dain:2013gma}. Consider a Newtonian, spherically 
symmetric object $\Omega$ with mass density $\mu$, mass $m$ and radius $R$, rotating with velocity $v$. Then its angular momentum is
\be
J=\int_{\Omega}\mu v\rho d^3x
\ee
where $\rho$ is the distance to the symmetry axis and $d^3x$ is the flat volume element. Since $\rho\leq R$ we obtain
\be
J\leq R\int_{\Omega}\mu v d^3x
\ee
Now we take from Einstein theory, the condition that the velocity should be smaller than the velocity of light, $v\leq 1$. 
This allows to bound the angular momentum as
\be\label{quasiobj}
J\leq R^2
\ee
where we have also used inequality \eqref{escape3} to bound the quasilocal mass $m=\int\mu\,d^3x$. Clearly, if instead of the trapped 
surface inequality \eqref{escape3} we use \eqref{escape2}, a factor $1/2$ would appear in \eqref{quasiobj}.

Finally, we consider a global inequality for Newtonian systems satisfying the condition $v\leq1$. We follow \cite{Anglada:2016dbu}. 
Let $\Omega$ be an ordinary object with mass density $\mu$, quasilocal mass $m$, characteristic radius $R$, equatorial radius $R_c$ 
(namely 
$R_c$ is the length, divided by $2\pi$, of the greatest axially symmetric circle on $\partial\Omega$) 
and angular momentum $J$. Then we expect that the total energy of the 
system is a sum of the gravitational and internal energies (included in the term $E_0$) and the rotational kinetic energy
\be
E\approx E_0+\frac{J^2}{2I}
\ee
where $I$ is the moment of inertia of the object. We bound the euclidean distance $\rho$ from the rotation axis, by the equatorial radius $R_c$, and use inequality \eqref{escape2}
\be\label{inertia}
I=\int_\Omega\mu\rho^2d^3x\leq mR_c^2\leq \frac{RR_c^2}{2}.
\ee
These considerations give
\be\label{globj}
E\gtrsim E_0+\frac{J^2}{RR_c^2}.
\ee
A factor $1/2$ would appear in \eqref{globj} if \eqref{escape3} was used instead of \eqref{escape2} in \eqref{inertia}.
Note that we obtain a lower bound on the total energy of the system in terms of $E_0$,  the angular momentum and two measures of size. One 
coming from the Hoop-like inequality \eqref{escape2} and the other coming from rotation. In other words, $R$ measures how localized matter 
is in $\Omega$ and $R_c$ measures how distributed matter is with respect to the rotation axis.

As opposed to \eqref{escape2}, \eqref{escape3} and \eqref{quasiobj}, which are quasilocal inequalities, \eqref{globj} is global in nature as it contains the 
total energy of the system.

What is remarkable is that these informal and naive arguments lead to similar inequalities that can be formally obtained from purely relativistic 
considerations about ordinary, non black hole objects. We review them in Sect.~\ref{sec:objects}.

\section{Basic definitions}
\label{sec:physical-quantities}

In order to extend inequalities \eqref{eq:penrose1}--\eqref{eq:13} to
dynamical black holes and  even to more general situations like ordinary
objects we need to introduce these elements with detail, and  properly define the physical quantities associated to them 
$(M,Q, J,A, \mbox{etc})$ in the
fully dynamical regime. 

Black holes are the main type of object we present in this article. That is the setting that originally motivated the study of 
the geometrical inequalities given in this article and where the most important results have been found so far. However, there are other quasilocal
objects that are relevant, namely isoperimetric surfaces and bounded regions representing ordinary objects. We present these objects in 
section  \ref{sec:BH}

On the other hand, the physical quantities studied in this review can be divided in three groups:
\emph{local quantities}, \emph{global quantities} and \emph{quasilocal
  quantities}.  Local quantities are tensor fields, global quantities are
associated to the whole spacetime and quasilocal quantities are associated with
finite regions. We define them in Sects.~\ref{sec:localq}, \ref{sec:globalq} and \ref{sec:quasilocalq} respectively.

Before proceeding further with these concepts, we fix the basic notation we use throughout the article.

Let $\M$ be a  4-dimensional manifold with metric $g_{\mu\nu}$ (with
signature $(-+++)$) and Levi-Civita connection $\nabla_\mu$. In the following,
Greek indices $\mu, \nu \cdots$ are always 4-dimensional. If, in addition, $g_{\mu\nu}$ satisfies Einstein equations 
\begin{equation}
  \label{eq:3a}
  G_{\mu\nu}\equiv  {^4R_{\mu\nu}-}\frac{1}{2}   {^4R} g_{\mu\nu}=8\pi T_{\mu\nu}-\Lambda g_{\mu\nu}\qquad \mbox{on } M,
\end{equation}
where $\Lambda$ is a cosmological constant, $T_{\mu\nu}$ is the energy momentum tensor, $^4R_{\mu\nu}$ is the Ricci tensor of the metric 
$g_{\mu\nu}$ and ${}^4R$, its scalar curvature, then we call $(M,g_{\mu\nu})$ a spacetime. Greek indices
$\mu, \nu,\ldots$ are 4-dimensional, they are raised and
lowered with the metric $ g_{\mu\nu}$ and its inverse $ g^{\mu\nu}$. 

Initial conditions for Einstein equations \eqref{eq:3a} are characterized by
an \emph{initial data set} $(\Si, h_{ij}, K_{ij}, \mu,
j^i)$ where $\Si$ is a connected 3-dimensional manifold, $h_{ij}
$ a (positive definite) Riemannian metric, $K_{ij}$ a symmetric
tensor field, $\mu$ a scalar field and $j^i$ a vector field on
$\Sigma$. These fields satisfy the constraint equations
\begin{align}
 \label{const1}
   D_j   K^{ij} -  D^i   K= -8\pi j^i\\
 \label{const2}
   {}^3R -  K_{ij}   K^{ij}+  K^2=16\pi \mu
\end{align}
on $\Si$. Here $D$ and ${}^3R$ are the Levi-Civita
connection and scalar curvature associated with $ {h}_{ij}$,
and $ K = K_{ij} h^{ij}$. Latin indices
$i,k,\ldots$ are 3-dimensional, they are raised and
lowered with the metric $ h_{ij}$ and its inverse $ h^{ij}$. 

\subsection{Asymptotically flat and cylindrical ends}
The initial data models an isolated system when the fields are weak far away from the
sources. This physical idea is captured in the following definition of
asymptotically flat initial data set. Let $B_R$ be a ball of finite radius $R$
in $\Rt$. The exterior region $U=\Rt\setminus B_R$ is called an \emph{end}.  On
$U$ we consider Cartesian coordinates $x^i$ with their associated euclidean
radius $r=\left( \sum_{i=1}^3 (x^i)^2 \right)^{1/2}$ and $\delta_{ij}$ to be
the euclidean metric components with respect to $x^i$.  A 3-dimensional
manifold $\Sigma$ is called \emph{Euclidean at infinity}, if there exists a compact
subset $\mathcal{K}$ of $\Sigma$ such that $\Sigma\setminus \mathcal{K}$ is the disjoint
union of a finite number of ends $U_k$.  The initial data set $(\Si, h_{ij},
K_{ij}, \mu, j^i)$ is called \emph{asymptotically flat} if $\Si$ is Euclidean
at infinity and at every end the metric $h_{ij}$ and the tensor $K_{ij}$
satisfy the following fall off conditions
\begin{equation}
  \label{eq:99}
  h_{ij}=\delta_{ij} +\hat h_{ij}, \quad K_{ij}=O(r^{-2}),
\end{equation}
where $\hat h_{ij}=O(r^{-1})$, $\partial_k \hat h_{ij}=O(r^{-2})$,
$\partial_l\partial_k\gamma_{ij}=O(r^{-3})$ and $\partial_k K_{ij}=O(r^{-3})$.
These conditions are written in terms of Cartesian coordinates $x^i$ attached
at every end $U_k$. Here $\partial_i$ denotes partial  derivatives with respect to these
coordinates.

The fall off conditions \eqref{eq:99} are far from being the minimal
requirements for the validity of the theorems presented in this article. We
have chosen these particular fall off conditions because they are simple to
present and they encompass a rich family of physical models. For more refined
assumptions we will refer to the original references.

See however, Sect.~\ref{sec:globalq} were the stronger fall off condition  \eqref{eq:30vv} on the second fundamental form $K_{ij}$
 is imposed. This  stronger requirement is necessary to make the integral in the definition of angular momentum converge.
 
An initial data may have more than one asymptotically flat end, and the asymptotic conditions \eqref{eq:99} should hold 
at each one of these ends. 

On the other hand, the initial data may have asymptotically cylindrical ends. They are defined in the following 
way, extracted from \cite{Chrusciel:2012nv, Chrusciel:2012np} (see also \citealp{dain10d}). An asymptotically cylindrical end of $\Si$ is $\mathbb R^+\times N$ where $N$ is a 
compact 2-manifold where $h_{ij}$ and $K_{ij}$ are conformal to fields having the asymptotic form
\be
\tilde h=dx^2+\tilde h^N+\mathcal O(e^{-\nu x}),\qquad \tilde K=\tilde K^N+\mathcal O(e^{-\nu x})
\ee
for some metric $h^N$ on $N$, a symmetric 2-tensor field $K^N$ on $N$, and a positive constant $\nu$.

\subsection{Black holes and other objects}\label{sec:BH}

\subsubsection{Black holes}
Black holes are global concepts referring to the causal structure of the whole spacetime and therefore, can not be defined in terms of local or quasilocal 
quantities. This property makes practical applications difficult to study and has led to the development of quasilocal meaningful characterizations 
of black holes. The intuitive idea that a black hole is a region of spacetime from which no signal can escape is captured by the notion of
trapped surface  described below. See the articles by \cite{Beig96c}, \cite{Dain03}, \cite{Booth:2005qc}, \cite{Jaramillo:2007km}, \cite{Mars:2009cj}, 
\cite{Hayward:2009kr}, and \cite{Senovilla:2011fk}, for further references, details and discussions on these 
quasilocal characterizations.

Consider an oriented spacetime $(M,g_{\mu\nu})$ and a closed, oriented,
spacelike 2-surface $\Su$ 
in $M$. Let $\ell^\mu$ and $k^\nu$ be the null vectors spanning the normal plane to $\Su$ and
normalized such that $\ell^\mu k_\mu = -1$ (note that there is a boost rescaling freedom
$\ell'^\mu =f \ell^\mu$, $k'^\mu = f^{-1} k^\mu$). In terms of $\ell^\mu$ and $k^\mu$, the induced metric and the
volume element on $\Su$ (written as spacetime projectors) are given by
$\gamma_{\mu\nu}=g_{\mu\nu}+\ell_\mu k_\nu+\ell_\nu k_\mu$ and
$\epsilon_{\mu\nu}=2^{-1}\epsilon_{\lambda\gamma\mu\nu}\ell^\lambda k^\gamma$
respectively. The expansions of the null congruences of geodesics with tangent vector fields $\ell^\mu$ and $k^\mu$ are 
\begin{equation}
  \label{eq:21}
  \theta_+:=\gamma^{\mu\nu}\nabla_\mu\ell_\nu, \quad \theta_-:=\gamma^{\mu\nu}\nabla_\mu k_\nu.
\end{equation}
The surface $\Su$ is called \textit{weakly (future) trapped} if $\theta_{\pm}\leq0$. The relevance of trapped surfaces comes from the singularity 
theorems of \cite{Penrose65} and \cite{Hawking71} (see also the review article by \cite{Senovilla:2014gza}). Under the Weak Cosmic Censorship Conjecture, future trapped surfaces in asymptotically flat initial data evolve
into black holes and therefore are fair quasilocal representatives of them. Moreover, the location of the trapped surface is related 
to the location of the event horizon. This, in particular, is important when analyzing size and shape of a trapped surface as a way to obtain 
information about the size and shape of a black hole.

In this article we are mainly concerned with two particular cases of trapped surfaces: 
\begin{itemize}
 \item \emph{marginally outer trapped surfaces} (MOTSs), for which $\theta_+=0$ on $\Su$,
\item  \textit{minimal} surfaces for which $\theta_\pm=0$. 
\end{itemize}

MOTS are typically located inside the event horizon in dynamical black hole spacetimes  and coincide with compact cross sections of the event 
horizon in stationary black hole spacetimes \citep{Andersson:2007gy}. There is, however an important point that one must keep in mind. This is, even when the trapped 
surfaces are inside the event horizon, their area need not be smaller than the black hole's area. 

Minimal surfaces have played a key role in the study of geometrical inequalities for black holes in General Relativity since the 
early days and especially since the proof of the positive mass theorem \citep{Huisken01} . Without mention of null expansions, minimal 
surfaces are characterized by the vanishing extrinsic curvature 
when seen as surfaces $\Su$ embedded in a 3-dimensional slice $\Si$. Note also that 
if the slice is part of a time symmetric initial data $(\Si, h_{ij}, K_{ij}\equiv0)$, then $\Su$ is also a MOTS. 
Moreover, as we show in Sect.~\ref{sec:proofAJ}, the variational characterization of stable minimal surfaces is closely related to the one for 
stable MOTSs.

With the concept of trapped regions, one can study a (globally hyperbolic) black hole spacetime as the manifold $(M=R\times\Si, \,g_{\mu\nu})$ where $\Si$ is a 
spacelike Cauchy surface with trapped inner boundary $\Su$. This is the approach taken to study quasilocal Lorentzian inequalities in section 
\ref{sec:QL}. Other alternative, mainly used in the study of global and quasilocal Riemannian inequalities is to consider $\Sigma$
as a surface with one asymptotically flat end (representing the region far away from the black holes), and as many extra ends as black holes
one wishes to consider \citep{Chrusciel:2007dd}. The extra ends will be asymptotically flat if the black holes are subextremal and 
asymptotically cylindrical if they
are extremal black holes. These extra ends are usually called punctures in the numerical community and this type of topology seems to be
more appropriate to numerical simulations than 3-surfaces with inner boundaries \citep{Immerman:2009ns}. So, in a sense, the black holes are
represented by non trivial topology on the initial surface $\Si$ (see \citealp{Gannon75, Lee76, Meeks82, Andersson:2007gy, Eichmair07, Andersson:2010jv, Chrusciel:2011iv}; and \citealp{Eichmair:2012jk}).

As an example, consider the Schwarzschild metric is standard coordinates
\be\label{sch1}
ds^2=-\left(1-\frac{2M}{r}\right)dt^2+\left(1-\frac{2M}{r}\right)^{-1}dr^2+r^2d\theta^2+r^2\sin^2\theta d\phi^2
\ee
This metric describes a black hole as the  manifold $M=R\times\Si$ with metric \eqref{sch1}, where $\Si$ is an asymptotically flat Riemannian manifold with 
inner boundary $\partial\Si=\{r=2M\}$. The inner boundary indicates the location of the event horizon. In this case, the sphere $r=2M$ 
is a minimal surface and a MOTS. There exists a coordinate system
where a doubling of $\Si$ is performed. Indeed, make the transformation to the isotropic coordinate $\bar r$ defined as
\be
\bar r:=\frac{1}{2}\left[r-M\pm\sqrt{r(r-2M)}\right].
\ee
Then the transformed metric
\be\label{sch2}
ds^2=-\left(\frac{1-M/2\bar r}{1+M/2\bar r}\right)dt^2+\left(1+\frac{M}{\bar r}\right)^{4}[d\bar r^2+\bar r^2d\theta^2+
\bar r^2\sin^2\theta d\phi^2]
\ee
is smooth on the doubled manifold. Note that $\bar r$ is double valued, it describes two copies of the exterior region of Schwarzschild. The horizon corresponds to
the surface $\bar r=M/2$. This metric is invariant under the inversion through the surface $\bar r=M/2$. The doubled Riemannian manifold
has two asymptotically flat ends at $\bar r\to 0$ and $\bar r\to\infty$  connected by a minimal surface at 
$\bar r=M/2$.
In this construction the presence of the black hole is manifested through the extra end at $\bar r\to0$.

\subsubsection{Isoperimetric surfaces and ordinary objects}

As we mention in the introduction, Sect.~\ref{sec:newt}, geometrical inequalities for non-black hole objects have gained impetus in 
recent years.

The first difficulty when studying these systems is the 
characterization of such ordinary objects. This problem does not appear in the black hole case where there is a well identified 
surface (the trapped surface) locating the black hole, to which one can associate  convenient stability properties.

Away from black holes, one can consider isoperimetric surfaces. These surfaces have  been studied within the context of geometrical 
inequalities in General Relativity, mainly in the context of Penrose inequality \citep{gibbons84, Malec:1992ap, Gibbons97, Gibbons06, corvino07}. Isoperimetric surfaces are such that its area is a critical 
point with respect to nearby surfaces enclosing a given volume. This variational characterization what makes them potentially useful for the study of inequalities.

Ordinary objects, on the other hand, are generally open, bounded sets in space where some specific matter fields have support. The main 
difficulty in this case is the lack of a variational characterization, which makes the obtention of geometrical 
inequalities hard to achieve. 
In Sect.~\ref{sec:ordinary} we describe what conditions have been imposed on the objects in order to produce the 
desired physical-geometrical estimate.

\subsection{Local physical quantities}\label{sec:localq}
The local physical quantities relevant for General Relativity are shown in Einstein equations \eqref{eq:3a} and, in particular, in 
Einstein constraints \eqref{const1}, \eqref{const2}. We focus in this section 
on the energy momentum tensor $T_{\mu\nu}$ describing matter fields.

It is useful to decompose  $T_{\mu\nu}$ into an electromagnetic part 
and a non-electromagnetic part
\begin{equation}
  \label{eq:14}
  T_{\mu\nu}=T^{EM}_{\mu\nu}+T^M_{\mu\nu},
\end{equation}
where $T^{EM}_{\mu\nu}$ is the electromagnetic energy momentum tensor given by
\begin{equation}
  \label{eq:4}
  T^{EM}_{\mu \nu}= \frac{1}{4\pi}\left(F_{\mu \lambda}
    F_\nu{}^{\lambda}-\frac{1}{4}g_{\mu \nu} F_{\lambda \gamma} F^{\lambda \gamma}  \right),
\end{equation}
and $F_{\mu\nu}$ is the (antisymmetric) electromagnetic field tensor
that satisfies Maxwell equations
\begin{align}
  \label{eq:maxwell}
  \nabla^\mu F_{\mu\nu} & =-4\pi j^{EM}_\nu, \\ 
\nabla_{[\mu} F_{\nu \alpha]} & =0.
\end{align}
where $ j^{EM}_\nu$ is the electromagnetic current.  

No specific matter model will be used, the only equation that $T_{\mu\nu}$ is
required to satisfy is the local conservation law \eqref{eq:16}

It is important to emphasize that, unless otherwise  stated, the tensors
$T^{EM}_{\mu\nu}$ and $T^M_{\mu\nu}$ are not, individually,  divergence free.

We assume that the matter fields satisfy  the \emph{dominant energy condition}, that is 
\begin{equation}
  \label{eq:15v}
  T_{\mu\nu} v^\mu w^\nu \geq 0,
\end{equation}
for all future-directed causal vectors $v^\mu$ and $w^\nu$.  We usually impose
this condition also on $T^M_{\mu\nu}$. 

Summarizing, the relevant local quantity is the energy momentum tensor
$T_{\mu\nu}$, which satisfies the local conditions \eqref{eq:16} and
\eqref{eq:15v}. Two important particular cases are  vacuum  $T_{\mu\nu}=0$ and
electrovacuum  $T^{M}_{\mu\nu}=0$.

\subsection{Global  physical quantities}\label{sec:globalq}

Global quantities are associated to isolated systems. An isolated system is an
idealization in physics that assumes that the sources are confined to a finite
region and the fields are weak far away from the sources. This kind of systems
are expected to have finite total energy, linear momentum, angular momentum and
charge.  In General Relativity there are several ways of defining isolated
systems. For our purpose the most appropriate definition is through initial
conditions for Einstein equations. Most results concerning global
inequalities discussed in this article has been proved, so far, in terms of
initial conditions.

Consider an initial data set $(\Si, h_{ij}, K_{ij}, \mu,
j^i)$ satisfying Einstein constraints \eqref{const1}, \eqref{const2}. If we take the initial data as a spacelike surface in the spacetime, with unit
timelike normal $t^\mu$, then the matter fields
$\mu$ and $j^i$ are given in terms of the energy momentum tensor by
\begin{equation}
  \label{eq:17x}
  \mu=T_{\mu\nu} t^\mu t^\nu, \quad j_i =T_{\mu i}t^\nu.
\end{equation}
The dominant energy condition (\ref{eq:15v}) implies
\begin{equation}
  \label{eq:65a}
  \mu^2 \geq j_ij^i.
\end{equation}

The decomposition (\ref{eq:14}) of the matter fields translates into
\begin{equation}
  \label{eq:5x}
  \mu=\mu_{EM}+\mu_M, \quad j_i=j_i^{EM}+j_i^M,
\end{equation}
where we have defined 
\begin{equation}
  \label{eq:10}
  \mu_{EM}=\frac{1}{4\pi}\left(E^iE_i+B^iB_i \right),\quad  j_i^{EM}=\epsilon_{ijk} E^jB^k,
\end{equation}
and the electric field $E$ and magnetic field $B$ are given by
\begin{equation}
  \label{eq:6x}
  E_\mu=F_{\mu\nu} t^\nu, \quad  B_\mu=- {}^*F_{\mu\nu} t^\nu,
\end{equation}
where the dual of $F_{\mu\nu}$ is defined with respect to the volume element
$\epsilon_{\mu\nu\lambda\gamma}$ of the metric $g_{\mu\nu}$ by the standard
formula
\begin{equation}
  \label{eq:dual}
  {}^*\alpha_{\mu_1\cdots \mu_{4-p}}=\frac{1}{p!}\alpha^{\nu_1\cdots \nu_p}
    \epsilon_{\nu_1\cdots \nu_p \mu_1\cdots  \mu_{4-p}}.
\end{equation}
The electric charge density $\rho_E$ is defined by
\begin{equation}
  \label{eq:29vc}
  D^i E_i=4\pi \rho_E.
\end{equation}
In vacuum we have $\mu=0$, $j^i=0$, and in electrovacuum,  $\mu_M=0$, $j^i_M=0$.

For asymptotically flat initial data the expressions for the total energy and
linear momentum of the spacetime were presented in \cite{Arnowitt62} (see also \citealp{Bartnik86} and \citealp{chrusciel86}) and they
are called the ADM energy and linear momentum. They are defined as integrals
over 2-spheres $\Su_r$ at infinity at every asymptotically flat end by the following formulae
\begin{align}
  \label{eq:EP}
  E &=\frac{1}{16\pi}\lim_{r\to \infty} \oint_{\Su_r} \left(\partial_j
    h_{ij}-\partial_i
    h_{jj}\right ) s^i \dsf \\
  P_i &= \frac{1}{8\pi} \lim_{r\to \infty} \oint_{\Su_r}
  \left(K_{ik}-Kh_{ik} \right ) s^k \dsf, \label{eq:EPP}
\end{align}
where $s^i$ is its exterior unit normal and $\dsf$ is the
surface element of the 2-sphere with respect to the euclidean
metric.   The energy $E$ and the linear momentum $P_i$ are components of
a 4-vector $(E,P_i)$.  The total mass of the spacetime is defined
by
\begin{equation}
  \label{eq:massadm}
  M=\sqrt{E^2-P^2},
\end{equation}
where we have used the notation $P^2=P_i P_j\delta^{ij}$.

Let $\beta^i$ be an infinitesimal generator for rotations with
respect to the flat metric associated with the asymptotically flat end $U$, then the angular
momentum $J$ in the direction of $\beta^i$ is given by
\begin{equation}
  \label{eq:35}
  J_\infty(\beta)=\frac{1}{8\pi} \lim_{r\to \infty} \oint_{\Su_r} (K_{ij}
  -Kh_{ij})\beta^i s^j  \dsf.
\end{equation}
The fall off conditions \eqref{eq:99} are not sufficient to ensure the
convergence of the integral \eqref{eq:35}, extra assumptions are needed. For
the results presented in this review which involve  the angular momentum
$J_\infty$, a stronger fall off condition on the second fundamental form
$K_{ij}$ is imposed
\begin{equation}
  \label{eq:30vv}
   K_{ij}=O(r^{-3}).
\end{equation}
In particular this assumption implies that  the linear momentum vanishes. 

The total electric and magnetic charges are given by \cite{Ashtekar00a}
\begin{equation}
  \label{eq:1x0}
  Q_{E\infty}=\frac{1}{4\pi} \lim_{r\to \infty} \oint_{\Su_r} E_i s^i \dsf. 
\end{equation}
\be
\label{eq:1xm}
  Q_{B\infty}=\frac{1}{4\pi} \lim_{r\to \infty} \oint_{\Su_r} B_i s^i \dsf 
\end{equation}
and we will usually denote them collectively as $Q_\infty$. We emphasize that for every asymptotically flat end $U_k$ we have the corresponding, in principle different, quantities
$E_{(k)}$, $ P^i_{(k)}$, $J^i_{ (k) \infty}$, $Q_{(k) \infty}$.

We use a subscript $\infty$ in the notation for $J_\infty$ and $Q_\infty$, to
distinguish them from the quasilocal quantities presented in Sect.~\ref{sec:quasilocalq}. However, since we will not discuss geometrical inequalities involving 
quasilocal mass or linear momentum we will not use a subscript $\infty$
in $E$ and $P^i$. We expect future extensions of inequalities 
\eqref{eq:penrose1}--\eqref{eq:13} in this direction, giving purely quasilocal geometrical inequalities. 

\subsection{Quasilocal  physical quantities}\label{sec:quasilocalq}

Quasilocal quantities depend on a bounded spacelike 3-dimensional region $\Sr$,
which can be thought as a subset of some initial data $\Sr\subset \Si$. There are
two kinds of quasilocal quantities, the first ones depend only on the boundary
of the region $\Omega$, that is a 2-dimensional spacelike closed surface that
we will denote by $\Su$. The second ones depend also on the interior of
$\Omega$ (for more details on this classification, that is both subtle and
important, see \citealp{Szabados04}). See also \cite{Wieland:2016dbc}, \cite{Chen:2013lza}, \cite{Epp:2013hua}, \cite{Tung:2009mh}, 
\cite{Yoon:2004hb}, and \cite{Nester:2004xm}). It turns out that for black holes only
quasilocal quantities of the first kind are relevant. For objects, quantities
of the second kind are also needed. For example, in spherical symmetry the
geodesic distance to the center is a relevant quasilocal measure of size. We
will present some of these measures with more detail in Sect.~\ref{sec:objects}.  Below, we concentrate on
quasilocal quantities that depend only on 2-dimensional closed spacelike
surfaces $\Su$.

On $\Su$ we define \emph{intrinsic}
and \emph{extrinsic} quantities. The former depend only on the induced
Riemannian 2-dimensional metric on the surface that we denote by
$q_{ij}$. The extrinsic quantities depend also on the extrinsic curvature of
the surface and possible additional fields like the electromagnetic fields.
The most important intrinsic quantity is the area $A(\Su)$ of the surface.  For
black holes, this is certainly the most relevant intrinsic quantity. But, even
for black holes, there exists also other intrinsic quantities that measure the
shape of the surface and satisfy geometrical inequalities. We will present them
in Sect.~\ref{sec:AJ}.

\subsubsection{Conserved quantities}\label{sec:cons}
For the discussion of conserved quasilocal quantities, we essentially follow Sect.~2 in
\cite{Szabados04} and \cite{Weinstein:1994bn}, see also \cite{Dain:2014xda}.

Consider an arbitrary energy-momentum tensor $T_{\mu\nu}$ which satisfies the
conservation equation 
\begin{equation}
  \label{eq:16}
  \nabla_\mu T^{\mu\nu}=0.
\end{equation}
on the curved background $(M,g_{\mu\nu})$
(we are not assuming Einstein equations \eqref{eq:3a}).  Assume that the
spacetime admits a Killing vector field $\eta^\mu$, that is
\begin{equation}
  \label{eqcm:12}
  \nabla_{(\mu} \eta_{\nu)}=0.
\end{equation}
For the present discussion, the vector $\eta^\mu$ is an arbitrary Killing field,
later on we will fix it to be the axial Killing field (i.e. the Killing field associated to axial symmetry). From equation
(\ref{eq:16}) and (\ref{eqcm:12}) we deduce that the vector
\begin{equation}
  \label{eqcm:5}
 Z^{\mu}= 8\pi T^{\mu \nu}\eta_\nu,
\end{equation}
is divergence free 
\begin{equation}
  \label{eq:64vv}
  \nabla_\mu  Z^{\mu}=0.
\end{equation}
The calculations involved in
the definitions of quasilocal quantities require  integration over domains with
different dimensions and the use of Stokes' theorem on them. Hence, it is
sometimes convenient to use differential forms instead of tensors in order to
highlight the geometrical meaning of the integrals.  In this section we denote them with
boldface. 

Let
$\boldsymbol{Z}$ be the 1-form defined by (\ref{eqcm:5}). Equation
(\ref{eq:64vv}) is equivalent to
 \begin{equation}
   \label{eq:63vv}
  d{}^*\boldsymbol{Z}=0.
 \end{equation}
Integrating (\ref{eq:63vv}) over an orientable, compact but otherwise arbitrary
4-dimensional region of the spacetime and using Stokes' theorem we obtain the
integral form of this conservation law. A particular relevant case is when the
4-dimensional region is a timelike cylinder such that its boundary is formed by
the bottom and the top spacelike surfaces $\Sr_1$ and $\Sr_2$ and the timelike
piece $\mathcal{C}$. For that case, we have
\begin{equation}
  \label{eq:29va}
   \int_{\Sr_2 }   {}^* \boldsymbol{Z}  -\int_{\Sr_1}   {}^*  \boldsymbol{Z} 
  =-\int_{\mathcal{C}}   {}^* \boldsymbol{Z}, 
\end{equation}
where the minus sign in the integral over $\Sr_1$ comes from the choice of the
normal.  The charge associated to $\boldsymbol{Z}$ of the 3-dimensional spacelike surface $\Sr$ is
defined by
\begin{equation}
  \label{eq:65}
Z(\Sr)=  \int_{\Sr}   {}^* \boldsymbol{Z},
\end{equation}
then we may write equation \eqref{eq:29va} as
\begin{equation}
  \label{eq:65z}
Z(\Sr_2)-Z(\Sr_1)=  -\int_{\mathcal C}   {}^* \boldsymbol{Z},
\end{equation}

Note that the quantities $Z(\Sr)$ are defined in terms of integrals over
3-dimensional spacelike surfaces. However, equation (\ref{eq:63vv}) implies, at
least locally, that there exists a 2-form ${}^*\boldsymbol{V}$ such that
\begin{equation}
  \label{eq:57}
 {}^*\boldsymbol{Z}=d {}^*\boldsymbol{V}.
\end{equation}
The 2-form $\boldsymbol{V}$ is called a superpotential for the 3-form $ {}^*\boldsymbol{Z}$.  We have chosen  the dual
${}^*\boldsymbol{V}$ instead of $\boldsymbol{V}$ in the definition
(\ref{eq:57}) in order to  make the analogy below, with the Maxwell form ${}^*\boldsymbol{F}$ more transparent. 
Then, using (\ref{eq:57}) and  Stokes' theorem once again we have 
\be
Z(\Omega)=\int_{\Sr}{}^*\boldsymbol Z=\int_{\Sr}d{}^*\boldsymbol V=\int_{\partial \Sr}{}^*\boldsymbol V 
\ee
Denoting by $\Su$ the boundary $\partial\Sr$ we arrive at the  conservation law
\begin{equation}
  \label{eq:60}
   Z(\Su_2)-Z(\Su_1)= -\int_{\mathcal{C}}   {}^*\boldsymbol{Z},
\end{equation}
where we have defined the quasilocal quantity $Z(\Su)$ by
\begin{equation}
  \label{eq:61}
 Z(\Su)= \int_{\Su} {}^* \boldsymbol{V}.
\end{equation}
For example, consider the
electromagnetic energy momentum tensor $T^{EM}$, and let $\eta^\mu$ be
a spacelike Killing vector (for instance, the axial Killing vector). Assume that $\Sr$ is
tangent to $\eta^\mu$. Then we have a
rotation axis $\beta$ and
\begin{align}
  \label{eq:67}
  Z^{EM}(\Sr)  & = 8\pi \int_{\Sr} T^{EM}_{\mu\nu} t^\mu \eta^\nu\\
&= 8\pi \int_{\Sr} E^j B^k \eta^i \epsilon_{ijk} \dv
\end{align}
where $t^\mu$ is the unit vector field normal to $\Omega$. In Minkowski, a rotation around an arbitrary axis $\beta^i$ is given by
\begin{equation}
  \label{eq:68}
  \eta_i =\epsilon_{ijk}\beta^j x^k,
\end{equation}
where $x^i$ are Cartesian coordinates. Then, the expression \eqref{eq:67} reduces to 
\begin{equation}
  \label{eq:69}
  Z^{EM}(\Sr) = 8\pi \int_{\Sr} E_{[i} B_{j]} x^i \beta^j. 
\end{equation}
This is the formula for the angular momentum (in the direction of $\beta$) of
the electromagnetic field used in textbooks (see, for example, \citealp{jackson99, zangwill2013modern}).

In Minkowski this construction provides, for each Killing vector field
$\eta^\mu$, the conservation law for all relevant physical  quantities
associated with the matter field $T^{\mu\nu}$
(i.e. energy, linear momentum, angular momentum).

\subsubsection{Electromagnetic charge}\label{sec:charge}

The most simple and important extrinsic quasilocal quantity on a closed 2-surface  $\Su$ is the
electromagnetic charge $Q(\Su)$. Its definition and properties serve as model for all
the other quasilocal quantities defined on $\Su$.   Let $\boldsymbol{F}$ be the 2-form corresponding to
the electromagnetic tensor $F_{\mu\nu}$, and let ${}^*\boldsymbol{F}$ be its
dual.  In terms of forms, Maxwell equations (\ref{eq:maxwell}) are written as
\begin{align}
  \label{eq:maxwellform}
  d {}^* \boldsymbol{F} &=4\pi {}^*\boldsymbol{j}^{EM},\\
d \mathbf{F} &=0. \label{eq:maxwellform2}
\end{align}

The conservation law for the current $\boldsymbol{j}^{EM}$ is obtained by taking an
exterior derivative to equation (\ref{eq:maxwellform}), namely
\begin{equation}
  \label{eq:19vv}
 d {}^*\boldsymbol{j}^{EM}=0. 
\end{equation}
Integrating (\ref{eq:19vv}) over a
4-dimensional timelike cylinder with boundaries $\Sr_1$, $\Sr_2$  and 
$\mathcal{C}$, as in Sect.~\ref{sec:cons}, gives
\begin{equation}
  \label{eq:29v}
   \int_{\Sr_2 }   {}^* \boldsymbol{j}^{EM}  -\int_{\Sr_1}   {}^*  \boldsymbol{j}^{EM} 
  =-\int_{\mathcal{C}}   {}^* \boldsymbol{j}^{EM}.
\end{equation}
The electric charge of the 3-dimensional spacelike surface $\Sr_2$ is
defined by
\begin{equation}
  \label{eq:65q}
Q_E(\Sr)=  \int_{\Sr}   {}^* \boldsymbol{j}^{EM}.
\end{equation}
Using equation (\ref{eq:maxwellform}) in the left hand side of
equation (\ref{eq:29v}) we can apply again Stokes' theorem over the 3-surfaces
$\Sr_1$ and $\Sr_2$ with boundaries $\Su_1$ and $\Su_2$ respectively. We obtain
\begin{equation}
  \label{eq:30v}
  Q_E(\Su_2)- Q_E(\Su_1)= -\int_{\mathcal{C}}   {}^*\boldsymbol{j}^{EM}, 
\end{equation}
where now  the electric charge $Q(\Su)$ is defined by the following surface integral
over $\Su$
\begin{equation}
  \label{eq:26}
  Q_E(\Su)=\frac{1}{4\pi}\int_{\Su} {}^* \boldsymbol{F}.
\end{equation}
Equation (\ref{eq:30v}) is the conservation law for the electric charge. It depends only on equation
(\ref{eq:19vv}). This equation implies that, at least locally, there exists a
2-form ${}^* \boldsymbol{F}$ such that (\ref{eq:maxwellform}) holds. 

In
electromagnetism, we start with the field equations
(\ref{eq:maxwellform})--(\ref{eq:maxwellform2}) and then we deduce
(\ref{eq:19vv}) and hence the conservation of $Q_E$. 

Similarly, by taking the exterior derivative of \eqref{eq:maxwellform2} one obtains that the magnetic charge
\begin{equation}
  \label{eq:26magn}
  Q_B(\Su)=\frac{1}{4\pi}\int_{\Su} \boldsymbol{F}
\end{equation}
is conserved \citep{Ashtekar00a},
that is
\be\label{consmagn}
 Q_B(\Su_1)= Q_B(\Su_2).
\ee
Equation \eqref{consmagn} means that $Q_B$ depends only
on the homology class of $\Su$. If $\Su$ can be shrunk to a point, then
$Q_B(\Su)=0$.

In particular, when $\boldsymbol{j}^{EM}=0$ in $\Omega$ the charge has the same value, namely
\begin{equation}
  \label{eq:62}
  Q_E(\Su_1)= Q_E(\Su_2). 
\end{equation}
That is, when no sources are present the electric charge $Q_E(\Su)$ also depends only
on the homology class of $\Su$.

In order to make contact between quantities written in terms of differential
forms and other equivalent expressions used in the literature written in terms
of tensors it is convenient to choose a tetrad adapted to a closed, oriented,
spacelike 2-surface $\Su$. Consider
null vectors $\ell^\mu$ and $k^\nu$ spanning the normal plane to $\Su$ and
normalized as $\ell^\mu k_\mu = -1$, leaving a (boost) rescaling freedom
$\ell'^\mu =f \ell^\mu$, $k'^\mu = f^{-1} k^\mu$. The induced metric and the
volume element on $\Su$ (written as spacetime projectors) are given by
$q_{\mu\nu}=g_{\mu\nu}+\ell_\mu k_\nu+\ell_\nu k_\mu$ and
$\epsilon_{\mu\nu}=2^{-1}\epsilon_{\lambda\gamma\mu\nu}\ell^\lambda k^\gamma$
respectively. The area measure on $\Su$ is denoted by $\ds$.

Using tensors and the adapted null vectors $\ell^\mu$ and $k^\mu$ defined
above, the electric and magnetic charges \eqref{eq:26}, \eqref{eq:26magn} are equivalent to
\begin{equation}
  \label{eq:41}
   Q_E(\Su)=\frac{1}{4\pi}\int_{\Su} F_{\mu\nu} \ell^\mu k^\nu \, \ds.
\end{equation}
\begin{equation}
  \label{eq:41magn}
   Q_B(\Su)=\frac{1}{4\pi}\int_{\Su} {}^*F_{\mu\nu} \ell^\mu k^\nu \, \ds.
\end{equation}

To relate quasilocal quantities with global quantities it is useful to
consider 2-surfaces $ \Su$ that are  boundaries of  compact subsets $\Sr$ of
the initial data $\Si$. Let $t^\mu$ denote the spacetime unit, timelike normal
of $\Sr$. Let $s^\mu$ be the unit, spacelike, normal of $\Su$ pointing in
the outward direction to $\Sr$ lying on  $\Sr$.  The outgoing and ingoing
null geodesics orthogonal to $\Su$ defined above are given by
$\ell^\mu=t^\mu+s^\mu$ and $k^\mu=t^\mu-s^\mu$.

The quasilocal electric and magnetic charges are given by the same expressions (\ref{eq:1x0}), \eqref{eq:1xm} but the
integrals are taken over the surface $\Su$, that is
\begin{equation}
  \label{eq:12}
  Q_E(\Su)=\frac{1}{4\pi}  \oint_{\Su} E_i s^i \ds. 
\end{equation} 
\be
 \label{eq:12magn}
  Q_B(\Su)=\frac{1}{4\pi}  \oint_{\Su} B_i s^i \ds. 
\end{equation} 
In particular, the total charge $Q_\infty$ is obtained as the limit
\begin{equation}
  \label{eq:9x}
  Q_\infty = \lim_{r\to \infty} Q(\Su_r),
\end{equation}
where the sequence of  surfaces $\Su_r$ are chosen  in the same asymptotic end.

\subsubsection{Angular momentum}

To define quasilocal angular momentum in general is a difficult problem (see
the review by \citealp{Szabados04}). However, for axially symmetric spacetimes there
exists a simple and well defined notion of quasilocal angular momentum which
was introduced by \cite{Komar59} (see also \citealp{Wald84}).  In the
literature this definition is usually discussed in vacuum settings, where the
angular momentum is conserved. Remarkably, it turns out that the quasilocal
inequalities of the form (\ref{eq:13}) are still valid in the non-vacuum case.
The inclusion of matter fields (in particular, electromagnetic fields) presents
some peculiarities in the definitions and also in the discussion of the
conservation (and non-conservation) of angular momentum.

We have seen in Sect.~\ref{sec:charge} that the superpotentials $\boldsymbol V$ for electric and magnetic charges are given by 
$\boldsymbol V_E=\boldsymbol F/4\pi$ and $\boldsymbol V_B={}^*\boldsymbol F/4\pi$.

However, there is in general no
explicit formula for the superpotential $\boldsymbol{V}_J$ that gives rise to angular momentum, in terms of the
fields. For example, consider the electromagnetic tensor $T^{EM}_{\mu\nu}$. Let
$A_\mu$ be the electromagnetic potential defined by
\begin{equation}
  \label{eq:28}
  F_{\mu\nu}= \nabla_\mu A_\nu -  \nabla_\nu A_\mu, \quad \boldsymbol{F} =d\boldsymbol{A}.
\end{equation}
Since $\boldsymbol{V}_J$ is calculated as an integral of $T^{EM}_{\mu\nu}$ (which
involves squares of $F_{\mu\nu}$), a naive counting of derivatives suggests
that $\boldsymbol{V}_J$ could be written as products between $F_{\mu\nu}$ and
$A_\mu$. However, it appears not to be possible to obtain such explicit
expression independent of the solutions \footnote{We thanks L. Szabados for
  clarifying this point}. In order to get such expression we need to impose the solution 
  to be symmetric with respect to the Killing vector  $\eta^\mu$ and also 
the surface of integration to be tangent to the Killing vector. In the following,
we will explicitly find the superpotential $\boldsymbol{V}_J$ for axially symmetric
solutions of Maxwell equation. We generalize the discussion presented in
\cite{dain12} to include a non-zero electromagnetic current
$\boldsymbol{j}^{EM}$ and also we make contact with other equivalent
expressions for the quasilocal angular momentum of the electromagnetic field
used in the literature.

Denote by $\eta^\mu$ the Killing field generator of the axial symmetry. The
orbits of $\eta^\mu$ are either points or circles. The set of point orbits
$\Gamma$ is called the axis of symmetry. Assuming that $\Gamma$ is a surface,
it can be proven that $\eta^\mu$ is spacelike in a neighborhood of $\Gamma$
(see \citealp{Mars:1992cm}). We will further assume that the Killing vector is
always spacelike outside $\Gamma$. Note that if this condition is not satisfied
then the spacetime will have closed causal curves, in particular it will not be
globally hyperbolic. The form $\eta_\mu$ will be denoted by
$\boldsymbol{\eta}$, and the square of its norm by $\eta$, namely
\begin{equation}
  \label{eq:7}
  \eta=\eta^\mu\eta_\mu=|\boldsymbol{\eta}|^2.
\end{equation}
We have used the notation $\eta^\mu$ to denote the Killing vector field and
$\eta$ to denote the square of its norm to be consistent with the
literature. However, to avoid confusions between $\eta^\mu$ and its square norm
$\eta$, we will denote the vector field $\eta^\mu$ by $\kif$ in equations
involving differential forms in the index free notation.

We assume that the Maxwell fields are axially symmetric, namely
\begin{equation}
  \label{eqcm:29}
  \pounds_\eta \boldsymbol{F}=0,
\end{equation}
where $\pounds$ denotes Lie derivative. 
Consider the 1-forms defined by 
\begin{equation}
  \label{eqcm:30}
  \boldsymbol{\alpha}=\boldsymbol{F}(\kif) , \quad  \boldsymbol{\beta}={}^*
  \boldsymbol{F}(\kif),
\end{equation}
where we have used the standard notation
$\boldsymbol{F}(\kif)=F_{\mu\nu}\eta^\mu$ to denote contractions of forms with
vector fields. From Maxwell equations
(\ref{eq:maxwellform})--(\ref{eq:maxwellform2}) and the condition
(\ref{eqcm:29}) we obtain
\begin{align}
  \label{eqcm:31}
  d \boldsymbol{\alpha} = 0, \quad d \boldsymbol{\beta}= -4\pi {}^*
  \boldsymbol{j}^{EM}(\bar\eta).
\end{align}
The first equation in \eqref{eqcm:31} implies that, locally, there exists a
function $\chi$ such that
\begin{equation}
  \label{eqcm:33}
 \boldsymbol{\alpha}=d \chi.  
\end{equation}

We calculate the 1-form $\boldsymbol Z$ defined in \eqref{eqcm:5} for the electromagnetic tensor $T^{EM}_{\mu\nu}$ where now $\eta^\mu$ is
the Killing vector field associated to axial symmetry. We denote it again  by $ \boldsymbol{Z}^{EM}$ and obtain
\begin{equation}
  \label{eqcm:16}
  \boldsymbol{Z}^{EM}  =2\left(\boldsymbol{F}(\alpha)  -\frac{1}{4}\boldsymbol{\eta}|F|^2\right).
\end{equation}
We want to write the integral of the 3-form ${}^*\boldsymbol{Z}^{EM}$ as a boundary
integral of a 2-form. In order to do that, we  use that $ \boldsymbol{\eta}
$ and $\boldsymbol{ \beta}$ satisfy the following identity
\begin{equation}
  \label{eqcm:17}
  {}^*(\boldsymbol{\eta} \wedge (\boldsymbol{F}(\alpha))=\boldsymbol{\alpha}
  \wedge 
  \boldsymbol{ \beta},
\end{equation}
and we  also use  the following general identity valid for arbitrary 1-forms
\begin{equation}
  \label{eqcm:36}
 {}^*(\boldsymbol{\eta} \wedge \boldsymbol{Z})\wedge \boldsymbol{\eta}=\eta 
{}^*\boldsymbol{Z} -  {}^*\boldsymbol{\eta} ( \boldsymbol{Z}(\kif)).
\end{equation}
Inserting \eqref{eqcm:17} and \eqref{eqcm:36} in \eqref{eqcm:16} we obtain
\begin{equation}
  \label{eq:50}
  {}^*\boldsymbol{Z}^{EM}=2\boldsymbol{\alpha} \wedge \boldsymbol{\beta} \wedge \boldsymbol{\hat \eta} +
  {}^*\boldsymbol{\hat\eta} (\boldsymbol{Z}(\bar \eta)),
\end{equation}
where the 1-form $\boldsymbol{\hat\eta}$ is defined by 
\begin{equation}
  \label{eqcm:38}
\boldsymbol{\hat\eta}=  \frac{\boldsymbol{\eta}}{\eta}.
\end{equation}
To write the first term in the right hand side of (\ref{eq:50}) as the exterior
derivative of a 2-form we use the following simple identity
\begin{equation}
  \label{eq:51}
  d(\chi\boldsymbol{\beta}\wedge\boldsymbol{\hat \eta})=\boldsymbol{\alpha}
  \wedge \boldsymbol{\beta} \wedge \boldsymbol{\hat \eta} +\chi
  d\boldsymbol{\beta}\wedge \boldsymbol{\hat \eta}  +\chi
  \boldsymbol{\beta}\wedge  d \boldsymbol{\hat \eta},
\end{equation}
where the potential $\chi$ is defined by (\ref{eqcm:33}). Putting
(\ref{eq:51}) in (\ref{eq:50}) we finally obtain
\begin{equation}
  \label{eq:52}
   {}^*\boldsymbol{Z}^{EM}=2  d(\chi\boldsymbol{\beta}\wedge\boldsymbol{\hat
     \eta}) +
8\pi \chi
\boldsymbol{\hat \eta} \wedge  {}^*\boldsymbol{j}^{EM}(\bar\eta) +\chi
  \boldsymbol{\beta}\wedge  d \boldsymbol{\hat \eta} + {}^*\boldsymbol{\hat\eta} (\boldsymbol{Z}(\bar \eta)),
\end{equation}
where we have used equation (\ref{eqcm:31}) to replace the term with
$d\boldsymbol{\beta}$ by $\boldsymbol{j}^{EM}$ in (\ref{eq:51}). 

We integrate equation (\ref{eq:52}) over a
3-surface $\Sr$ tangential to $\eta^\mu$, with boundary $\Su$. The third and
the fourth term  in (\ref{eq:52}) do not contribute to the integral because
\begin{equation}
  \label{eqcm:39}
  d \boldsymbol{\hat\eta}(\kif)=0, \quad  \boldsymbol{\beta}(\kif)=0,
\end{equation}
and also the restriction
of the 3-form ${}^*\boldsymbol{\hat\eta}$ to $\Sr$ is zero.  
Hence, we obtain the final result
\begin{equation}
  \label{eq:54}
\frac{1}{8\pi}  \int_{\Sr}  {}^*\boldsymbol{Z}^{EM}= -J_{EM}(\Su) + 
\int_{\Sr} \chi
\boldsymbol{\hat \eta} \wedge  {}^*\boldsymbol{j}^{EM}(\bar\eta),
\end{equation}
where we have defined the  quasilocal angular momentum of the electromagnetic
field $J_{EM}(\Su)$   by
\begin{equation}
  \label{eq:21v}
  J_{EM}(\Su)=- \frac{1}{4\pi} \int_\Su \chi\boldsymbol{\beta}  \wedge \boldsymbol{\hat\eta}.
\end{equation}
The 2-form $ \chi\boldsymbol{\beta}  \wedge \boldsymbol{\hat\eta}$ is the
superpotential ${}^*\boldsymbol{V}_J$ used  in \eqref{eq:61}. We remark, however, that the expression \eqref{eq:21v} is valid only for 
axially symmetric solutions
(i.e. we have assumed \eqref{eqcm:29}) and axially symmetric surfaces (i.e. the
Killing field $\eta^\mu$ is tangent to $\Su$). 

Note that equation \eqref{eq:54} is valid for non-zero sources $\boldsymbol{j}^{EM}$.  We discuss the case $\boldsymbol{j}^{EM}=0$, 
and more generally, the case $\boldsymbol{j}^{EM}(\bar\eta)=0$ for axially symmetric initial data in Sects.~\ref{sec:proofMJ} and 
\ref{sec:proofAJ}.

To write the angular momentum \eqref{eq:21v} in terms of the potential $A_\mu$ defined by \eqref{eq:28} we use that
\begin{equation}
  \label{eq:33}
  \chi = A_\mu \eta^\mu,
\end{equation}
where we have assumed that the potential $A_\mu$ is also axially symmetric, that is
\begin{equation}
  \label{eq:34x}
   \pounds_\eta \boldsymbol{A}=0.
\end{equation}
Then inserting (\ref{eq:33}) in (\ref{eq:21v})  we get
\begin{equation}
  \label{eq:39}
  J_{EM}(\Su)= \frac{1}{4\pi} \int_\Su (A_\lambda \eta^\lambda) F_{\mu\nu} l^\mu k^\nu \, \ds. 
\end{equation}

We consider now the angular momentum of the gravitational field.  The analog of
the electromagnetic current 1-form $\boldsymbol{j}^{EM}_\mu$ is played by the
1-form $\boldsymbol{\hat{Z}}$ defined by
\begin{equation}
  \label{eqcm:14}
\boldsymbol{\hat{Z}}\equiv  \hat{Z}_\mu ={}^4R_{\mu\nu}\eta^\nu.
\end{equation}
Using  the Killing equation for $\eta^\mu$ we obtain
\begin{equation}
  \label{eq:64}
  \nabla_\mu  \hat{Z}^{\mu}=0.
\end{equation}
This equation is equivalent to
 \begin{equation}
   \label{eq:63}
  d{}^*\boldsymbol{\hat{Z}}=0.
 \end{equation}
Hence, we have conservation law for $\boldsymbol{\hat{Z}}$, as we have discussed above for the matter fields and the form 
$\boldsymbol{Z}$ defined by (\ref{eqcm:5}). For $\boldsymbol{\hat{Z}}$, the Komar identity given by 
 (see \citealp{Wald84}) 
\begin{equation}
  \label{eqcm:13}
  d  {}^*d \boldsymbol{\eta}=  2 {}^* \boldsymbol{\hat{Z}},
\end{equation}
provides an explicit formula for the superpotential of $\boldsymbol{\hat{Z}}$. The quasilocal Komar angular momentum is defined by  
\begin{equation}
  \label{eqcm:1}
  J_K(\Su)= \frac{1}{16\pi}
  \int_\Su {}^* d \boldsymbol{\eta},
\end{equation}
where $\Su$ is an arbitrary spacelike closed 2-surface.  The conservation law for angular momentum in axial
symmetry, which is the exact analog to the charge conservation, reads
\begin{equation}
  \label{eqcm:41}
J(\Su_1)- J(\Su_2)=\frac{1}{8\pi} \int_{\mathcal{C}}   {}^* \boldsymbol{\hat{Z}}. 
\end{equation}
The right hand side of this equation represents the change in the angular
momentum of the gravitational field which is produced by the left hand side,
namely the angular momentum of the matter fields.

As in the case of the charge, integrating on the spacelike domain $\Sr$ with boundaries $\Su_1$ and $\Su_2$, gives
\begin{equation}
  \label{eq:59a}
   J_K(\Su_2)- J_K(\Su_1)= -\int_{\Sr}   {}^* \boldsymbol{\hat{Z}}. 
\end{equation}
In particular, in vacuum we have $\boldsymbol{\hat{Z}}=0$ and hence the angular momentum  has the same value on both surfaces, namely
\begin{equation}
  \label{eq:62a}
  J_K(\Su_1)= J_K(\Su_2). 
\end{equation}
That is, for axially symmetric vacuum solutions of Einstein equations the angular momentum  $J_K(\Su)$ depends only
on the homology class of $\Su$. If $\Su$ can be shrunk to a point, then
$J_k(\Su)=0$.

In terms of tensors we have the equivalent expression for the Komar angular momentum
\begin{equation}
  \label{eq:42}
   J_K(\Su)=\frac{1}{16\pi}\int_\Su
  \epsilon_{\mu\nu\lambda\gamma}\nabla^\lambda \eta^\gamma\, \ds.
\end{equation}

Using Einstein equations (\ref{eq:3a})   we can relate the form $\boldsymbol{Z}$ of the matter fields (\ref{eqcm:5}) and the form 
$\boldsymbol{\hat{Z}}$ of the gravitational field 
\begin{equation}
  \label{eqcm:15}
   \hat Z_\mu =8\pi (T_{\mu\nu}\eta^\nu-\frac{1}{2}T \eta_\mu) -\Lambda \eta_\mu,
\end{equation} 
that is
\begin{equation}
  \label{eq:66}
  \boldsymbol{\hat{Z}}= \boldsymbol{Z}-(4\pi T + \Lambda)\boldsymbol{\eta}. 
\end{equation}

According to (\ref{eq:14}) and (\ref{eq:3a}) we decompose  
$\boldsymbol{Z}$   into the electromagnetic part and 
the non-electromagnetic part 
\begin{equation}
  \label{eq:53}
  \boldsymbol{Z}=\boldsymbol{Z}^{EM}+\boldsymbol{Z}^{M}. 
\end{equation}

The total angular momentum is defined by
\begin{equation}
  \label{eq:43}
  J_T(\Su)= J_{EM}(\Su) +J_{K}(\Su).
\end{equation}
It satisfies the conservation law
\begin{equation}
  \label{eq:36}
    J_T(\Su)= \frac{1}{8\pi}  \int_{\Sr}  {}^*\boldsymbol{Z}^{M} +\chi
\boldsymbol{\hat \eta} \wedge  {}^*\boldsymbol{j}^{EM}(\bar\eta).
\end{equation}
Note that the cosmological constant term does not contribute because the surface
is tangential to $\eta^\mu$. In terms of tensors,  (\ref{eq:36}) is
written as
\begin{equation}
  \label{eq:70}
  J_T(\Su)= \int_{\Sr} T^{M}_{\mu\nu} t^\mu \eta^\nu + A^\mu\eta_\mu j^{EM}_\nu
  t^\nu  \, \dv 
\end{equation}

There exists a very simple  expression for the Komar integral on an
initial data, namely
\begin{equation}
  \label{eq:38}
  J_K(\Su)=\frac{1}{8\pi}\int_\Su K_{ij} \eta^i s^j \ds.
\end{equation}
As in the case of the electric charge, this expression is the quasilocal version
of the global expression (\ref{eq:35}) (recall that $s_i\eta^i=0$ near
infinity).  In particular, assuming the fall off conditions, we have
\begin{equation}
  \label{eq:11}
  J_\infty(\eta)= \lim_{r\to \infty} J_K(\Su_r),
\end{equation}
and
\begin{equation}
  \label{eq:44}
  \lim_{r\to \infty} J_{EM}(\Su_r)=0.
\end{equation}

\section{Global inequalities for black holes}
\label{sec:global}

In this section we present results involving the total mass of a black hole and some of its quasilocal physical parameters like the electromagnetic
charge and angular momentum. We group the results into two sections. The first group, in Sect.~\ref{sec:MQ} refers to inequalities involving
total mass and electromagnetic charge, with zero angular momentum.  The second group incorporates angular momentum. As we discuss in section 
\ref{sec:physical-quantities}, in order to have a well defined conserved angular momentum, axial symmetry is required, whereas the pure charge case
needs no symmetry at all to be well formulated. This difference makes the techniques used to solve the problems very different as we discuss
below. 
Before going into the details, we want to make two remarks. 

\textit{Settings.} The geometrical inequalities presented in this section are proven for a set of initial data $(\Si, h_{ij}, K_{ij})$ for Einstein
equations. Evolution arguments are not considered. Moreover, Einstein constraint equations are used in a crucial way. The initial
surface has an asymptotically flat end, where the ADM is computed. Also, it may have an inner boundary, connected or not,  or be 
complete, with at least one more end. These features capture the presence of the black hole (see Sect.~\ref{sec:BH}). In the introduction, 
Sect.~\ref{sec:phys-heur-argum}, we give arguments indicating that the global black hole inequalities are valid for 
all times if they are valid at some initial time.

\textit{Inequality producer.} We wish to mention here  the features in the systems considered that ultimately produce the
inequalities.  Note that we are not thinking about the general qualities of systems in the theory that allow such inequalities (this 
was discussed at the beginning of the introduction). But we wonder about the underlying mathematical hypothesis or condition on the 
initial data $(\Si, h_{ij}, K_{ij})$ that translate 
into the desired relation between the physical and geometrical 
parameters. Interestingly, in order to  derive the global inequalities presented in this section, the only requirement
are certain energy conditions and
the presence of at least two ends on $\Si$ if $\Sigma$ has no inner boundary (one of which needs to be asymptotically flat 
to have a well defined ADM mass). 
However, as we present in Sect.~\ref{sec:MQ}, the Mass-Charge inequality is also proven when $\Sigma$ is an asymptotically
flat manifold with inner trapped boundary. This is in contrast with the Penrose inequality \citep{Mars:2009cj}, which is also global, but 
where the area of a closed 2-surface is included explicitly into the estimate
and therefore, extra assumptions on such 2-surface must be imposed.

\subsection{Mass-Charge}
\label{sec:MQ}

The Mass-Charge inequality arises as a way to refine the positive mass theorem \citep{Schoen79b, Schoen81c, witten81} and to give a strictly 
positive lower
bound for the total mass in the Einstein--Maxwell theory. 

\subsubsection{Results}\label{sec:resultsMQ}
Below is the main theorem showing this result. Its proof is greatly due to \cite{Gibbons82}, and to \cite{Gibbons83}, but many authors have contributed to the final version. We describe their particular input after the statement.

\begin{theorem}\label{thQM}
Let $(\Si, h_{ij}, K_{ij}, E^i, B^i, \mu_M, j^i_M)$ be a strongly asymptotically flat initial data for the Einstein-Maxwell equations, 
with $(\Si, h_{ij})$ complete or having a weakly outer trapped inner boundary. Assume that matter fields satisfy the energy condition
\begin{equation}
  \label{eq:8}
 \rho_{EM}\leq  \sqrt{\mu_M^2-|j_M|^2}.
\end{equation}
Then, on every end 
\begin{equation}
  \label{eq:3}
  |Q_\infty| \leq M ,
\end{equation}
where $M$ and $Q_\infty=\sqrt{Q_{E,\infty}^2+Q_{B,\infty}^2}$ are computed at the same end. 
Moreover
\begin{enumerate}[i)]
 \item \label{rigidQM1}if the equality in (\ref{eq:3}) is
attained then the associated spacetime is, locally, an Israel-Wilson-Perj\'es
metric.
\item \label{rigidQM2}if the initial data
  are maximal (i.e. $K=0$) and electro-vacuum, then the equality in (\ref{eq:3})
  holds if and only if the data set arises from the Majumdar-Papapetrou
  spacetime.
  \item \label{rigidQM3} if the Dirac-Jang equations have an appropriate solution then the equality in (\ref{eq:3})
  holds if and only if the data set arises from the Majumdar-Papapetrou
  spacetime.
\end{enumerate}

\end{theorem}

As we mention above, \eqref{eq:3} is proven in \cite{Gibbons82, Gibbons83}. They use spinorial arguments similar to ones 
used in Witten's proof of the
positive mass theorem \citep{witten81}. \cite{Gibbons82} shows that the equality in \eqref{eq:3} holds if and only if there
exists a super covariantly constant spinor, and the Israel--Wilson--Perj\'es and  Majumdar--Papapetrou metrics are discussed in this context.

\cite{Tod:1983pm} (see also \citealp{Herzlich199897} and \citealp{Horowitz84}) addresses the rigidity statement \eqref{rigidQM1} by finding all smooth spacetimes admitting such super 
covariantly constant spinors. They are gravitational and electromagnetic 
plane waves possibly with dust, and metrics describing charged rotating dust. Particular cases of the latter are the Israel--Wilson--Perj\'es
metrics, some of the Bonnor metrics and the Majumdar--Papapetrou metric.

\cite{Chrusciel:2005ve} prove the rigidity statement \eqref{rigidQM2} by showing that under 
certain conditions, the Israel--Wilson--Perj\'es metrics are of Majumdar--Papapetrou class.

\cite{Bartnik05} generalize the proof of \eqref{eq:3} to
include low differentiable metrics, namely $h_{ij}\in H^2_{loc}$ and $K_{ij}, E^i, B^i \in H^1_{loc}$. However, the equality is left 
open, as the
classification of metrics admitting super covariantly constant spinors of \cite{Tod:1983pm} does not apply in the rigidity case. 
  
The rigidity statement  \eqref{rigidQM3} is proven by \cite{Khuri:2013kaa} assuming that a system of equations, namely the
 Dirac--Jang equations, has appropriate solutions. They also assume what they call \textit{the charged dominant condition}, which reads 
 $\rho_{EM}\leq \mu_M+|j_M|$ and is stronger than condition \eqref{eq:8}.

 A related Mass-Charge inequality was
proved by \cite{Moreschi84} with similar spinorial techniques, and also later by \cite{Bartnik05}. More precisely, if instead of \eqref{eq:8}, matter fields
satisfy 
\be\label{eq:moreschi}
\alpha\rho_{EM}\leq  \sqrt{\mu_M^2-|j_M|^2}
\ee
for some $\alpha\in(0,1]$, then the inequality $|Q_\infty| \leq \alpha M$ holds. Note that when $\alpha=1$ it reduces to \eqref{eq:8}.
This result is relevant for ordinary matter (see the first remark below).

\subsubsection{Discussion}\label{sec:discussionMQ}
 A few comments about the hypotheses and applications follow.

 The energy condition (\ref{eq:8}) can be interpreted as a
local version of the global inequality (\ref{eq:3}) . Namely, write  (\ref{eq:3}) as
\begin{equation}
  \label{eq:37}
 \sqrt{P^2+ Q^2_\infty}\leq E,
\end{equation}
then $P$, $Q_\infty$ and $E$ are the global quantities corresponding to the local
quantities $j_M$, $\rho_E$ and $\mu_M$ in (\ref{eq:8}). In that sense,
condition (\ref{eq:8}) looks rather natural. However it is important to recall
that ordinary charged matter can  violate this condition and, of course,
also the global inequality (\ref{eq:3}) (see the discussion in \citealp{Gibbons82, Horowitz84, Moreschi84} and \citealp{dain12}). The ultimate reason for
that is that the mass charge relation of the electron violates the inequality
(\ref{eq:3}) for several orders of magnitude.  In fact, the local condition (\ref{eq:8})
allows only matter fields with very small amount of electric charge.  One way to avoid this limitation is to relax the condition on 
matter fields, for instance, by asking \eqref{eq:moreschi}. Then the inequality obtained has a much wider applicability. 
In the
electro-vacuum case we have $\rho_E=0$, then condition (\ref{eq:8}) reduces to the
dominant energy condition for the non electromagnetic matter fields.

 The manifold $\Si$ can have multiple asymptotic ends, on different ends the
value of the quantities $E,P, Q_\infty$ are different and the inequality
(\ref{eq:3}) holds on every end.  Also, the theorem admits a manifold with an
inner boundary given by the 2-surface $\Su$. Consider the simplest case, where the manifold
$\Si$ is $\Rt$ (which, of course, means that it has only one asymptotic end and
no inner boundary). For that case, in order to have non-zero $Q_\infty$ we need
to have charged matter in the interior, that is $\rho_E \neq 0$. The value of
$Q_\infty$ represents the total charge of the spacetime, it will be in general
different than the value of the charge calculated for an arbitrary surface
$Q(\Su)$ in the interior.  An important spacetime that satisfies these
conditions is the electrically counterpoised dust studied by Bonnor (see
\citealp{Bonnor:1980dj, bonnor98} and reference therein). These are explicit
solutions that describe static configurations of charged dust with
$\mu_M=|\rho_E|$, and hence the gravitational attraction is exactly balanced by
the electric repulsion. The shape of the configuration can be arbitrary and does  not need to have
any spatial symmetry.  These spacetimes achieve the equality $M=|Q_\infty|$ in
(\ref{eq:3}).

If we assume electro-vacuum, in order to have a non-zero $Q_\infty$ we need to
allow either a non-trivial topology with multiple asymptotic ends or a
non-trivial inner boundary $\Su$.  The Reissner--Nordstr\"om black hole initial
data are the model example of both cases: either we consider them as a complete
manifold with two asymptotically flat ends or as a manifold with inner boundary
$\Su$ at the black hole horizon which is weakly outer trapped.  The equality
$M=|Q_\infty|$ is achieved by the extreme Reissner--Nordstr\"om black hole, which has
one asymptotically
flat end and one cylindrical end. On the other hand, initial data of the super
extreme Reissner--Nordstr\"om metric does not satisfies the hypotheses since the
Riemannian manifold $(\Si,h_{ij})$ is not complete and does not have any weakly
outer trapped 2-surface.

The Majumdar-Papapetrou spacetime also satisfies the hypotheses of theorem
\ref{thQM}. The initial data has only one asymptotically flat end and an arbitrary
number of extra cylindrical ends.  This spacetime describes the equilibrium
configuration of multiple extreme black holes for which the gravitational
attraction is balanced by the electric repulsion. It satisfies $M=|Q_\infty|$ (see \citealp{Hartle:1972ya} for a 
discussion of these spacetimes). The extreme Reissner--Nordstr\"om black hole is
a particular case of Majumdar-Papapetrou with only one cylindrical end.

The Israel--Wilson--Perj\'es metrics are characterized
by the existence of a `super-covariantly constant' spinor field (for details
about this metrics see \citealp{Tod:1983pm, Chrusciel:2005ve} and references
therein). Example of this class of metrics are the above mentioned electrically
counterpoised dust and the Majumdar--Papapetrou metrics. We emphasize that the
equality in (\ref{eq:3}) can be achieved by a non electro-vacuum
solution. However, for the electro-vacuum case a stronger result is given in statement \eqref{rigidQM2}.

\subsection{Mass-Angular Momentum-Charge}
\label{subsec:MJ}
The inclusion of the angular momentum in the inequality \eqref{eq:24i} involves
completely different techniques as the one used in theorem \ref{thQM}.  In
particular no spinorial proof of this inequality with angular momentum is
available so far (see however \citealp{Zhang99} where a related inequality is
proven using spinors).

\subsubsection{Results}\label{sec:resultsMJ}
Below is the Mass-Angular Momentum-Charge theorem, the proof in vacuum is mainly due to \cite{Dain06c}. Some of the later generalizations and 
refinements by other authors are included in the statement and discussed after it.

\begin{theorem}
\label{t:main-1}
Let $(\Si, h_{ij}, K_{ij}, E^i, B^i)$ be  an electrovacuum, axially symmetric, maximal, initial data set with
two asymptotic ends. One end is asymptotically flat, where the
fall off condition \eqref{eq:30} is assumed for the second fundamental form. The other end is asymptotically flat or cylindrical.

Then, the following inequality holds
at the  asymptotically flat end
\begin{equation}
  \label{eq:42b}
  \frac{Q_\infty^2+\sqrt{Q_\infty^4+4J_\infty^2}}{2} \leq M^2.
\end{equation}
The equality in (\ref{eq:42b}) holds if and only if the initial data
corresponds to the extreme Kerr--Newman black hole. 
\end{theorem}

The proof of the inequality \eqref{eq:42b}
was
provided by Dain in a series of articles \citep{Dain05c, Dain05d,
Dain05e}, which end up in the global proof given in
\cite{Dain06c}. There it is shown that  \eqref{eq:42b} holds in vacuum and for a class of axially symmetric black hole initial data 
known as Brill data. The argument exploits the relation between a certain mass functional $\mM(\eta,\omega)$  and the energy for 
harmonic maps from $\Rt$ to the Hyperbolic plane \citep{Carter73, Ernst68} ($\eta$ and $\omega$ are the 
norm and twist potential of the axial Killing vector). See Sect.~\ref{sec:proofMJ} for the definition and properties of $\mM$ and details
about the proof of theorem \ref{t:main-1}.

Along the same lines, \cite{Chrusciel:2007dd, Chrusciel:2007ak} give a simpler proof of the theorem and avoid some technical assumptions on Dain's proof but
consider  a class  of axially symmetric initial data that do not contain the limit case in vacuum, namely,  extreme Kerr. They allow 
positive matter density that does not enter explicitly into the inequality. Moreover, they assume the existence of a twist potential  
$\omega $ which is known to exist in vacuum.

Electromagnetic charge is brought into the inequality by \cite{Chrusciel:2009ki} and \cite{Costa:2009hn}.
In particular, in \cite{Costa:2009hn} an appropriate potential, related to the twist potential for the axial Killing vector is shown to 
exist. This allows the definition of a new mass function $\mM(\eta,\omega,\psi,\chi)$, where $\eta$ and $\omega$ are as before and
$\psi$, $\chi$ are the electromagnetic potentials (see Sect.~\ref{sec:proofMJ}). As in \cite{Chrusciel:2007ak}, the initial
data considered does not include extreme Kerr--Newman and therefore, the rigidity statement can not be analyzed.

\cite{Schoen:2012nh} make
several improvements on the assumptions, they give a relevant lower bound for
the difference between general data and the extreme Kerr–Newman data and they
prove the rigidity statement in the charged case. 

In theorem \ref{t:main-1} the data are assumed to be maximal.  For the case
with no charge, the maximal condition is relaxed to a small trace case
assumption by \cite{zhou12}. 

Maximality is completely removed
by \cite{Cha:2014oja, Cha:2014yja}, including also non electromagnetic matter fields $\mu_M,j_M$ that satisfy the charged dominant energy condition 
$\mu_M\geq|j_M|$. Instead, it is assumed that a system of equations has
appropriate solutions.  In these articles a deformation procedure of the
initial data is constructed that provides a natural and clean way to
automatically generalize geometrical inequalities proved in the maximal case.

\subsubsection{Discussion}\label{sec:discussionMJ}
Before presenting related results to theorem \ref{t:main-1}, a few remarks are in order.

We begin by comparing  theorem \ref{t:main-1}  and the pure charge theorem, \ref{thQM}.  In theorem
\ref{t:main-1} axial symmetry is assumed globally on the initial data, in
contrast with theorem \ref{thQM} where there are no symmetry assumptions.  As
we have seen in Sect.~\ref{sec:phys-heur-argum}, on physical grounds, the
inequality \eqref{eq:42b} is not expected to hold for non-axially symmetric
data. General families of non-axially symmetric counter examples of the
inequality (\ref{eq:42b}) have been constructed by 
\cite{huang11} for pure vacuum and complete manifolds.

In theorem \ref{t:main-1} electrovacuum is assumed.  It is conceivable that,
using similar techniques as in the current proof of this theorem, the
electrovacuum assumption can be slightly relaxed by assuming, in analogy with
the assumption \eqref{eq:8} in theorem \ref{thQM}, that the matter fields have
small angular momentum and charge. We expect that this assumption will be
rather unphysical, since ordinary rotating matter can easily violate the
inequality (\ref{eq:42b}). It is however interesting that in theorem \ref{thQM} a
rigidity statement is obtained even in the non-electrovacuum case. It is not
known whether an analogous rigidity result holds for axially symmetric matter
fields with non trivial angular momentum in which the equality in
(\ref{eq:42b}) is achieved. Note however, that the rigidity statement in theorem
\ref{thQM} depends strongly on the spinorial proof.  It appears to be unlikely
that a generalization of this rigidity result holds for the case of angular
momentum.

On the other hand, we expect that the dominant energy condition should be required, as there are numerical examples by \cite{Bode:2011xz} of spinning black holes with matter violating the null (and dominant) energy condition, with $J\geq M^2$.

No inner boundary is allowed in theorem \ref{t:main-1}. The inclusion of an inner
boundary (which presumably should be a weakly trapped surface as in theorem
\ref{thQM}) appears to be a difficult and a relevant problem. An inner boundary
requires appropriate boundary conditions for the variational problem used in
the proof of theorem \ref{t:main-1}.  The results presented by 
\cite{Gibbons06} and \cite{Chrusciel:2011eu}
contribute in this direction, but so far the problem remains open.

Since electrovacuum is assumed, in order to have non-zero charge and angular
momentum the manifold should have a non-trivial topology. In theorem
\ref{t:main-1} a particular geometry is assumed: manifolds with two 
ends. This is certainly the stronger restriction of this theorem.
Let us discuss this point in detail.

The model initial data set that satisfies all the hypotheses of theorem
\ref{t:main-1} is a slice $t=constant$ in the standard Boyer-Lindquist
coordinates of the Kerr--Newman black hole.  In the non-extreme case these
initial data have two asymptotically flat ends, where the standard fall off
conditions (\ref{eq:99}) are satisfied, plus the stronger fall off condition
(\ref{eq:30vv}) of the second fundamental form. However, in the extreme case,
the geometry of the initial data changes: one end is cylindrical and the other
is asymptotically flat. That is, in order to include the extreme case more
general fall off conditions need to be allowed on the one of the ends.

\subsubsection{Multiple black holes}

For multiple ends the problem is
open. There exist however the following very interesting  results.  In
order to describe them, we need to highlight some properties of the mass functional $\mf$ (see Sect.~\ref{sec:proofMJ} for more details). This functional
represents a lower bound for the mass. Moreover, the global minimum of this
functional (under appropriate boundary conditions which preserve the angular
momentum) is achieved by a harmonic map with prescribed singularities.  This
is the main strategy in the proofs of all the previous theorems which are valid
for two asymptotic ends. Remarkably enough,
\cite{Chrusciel:2007ak} prove the existence and uniqueness of this singular
harmonic map for manifolds with an arbitrary number of asymptotic ends, and
then, as a corollary they prove the following result

\begin{theorem}
  \label{Piotr-Gilbert}
  Consider an axially symmetric, vacuum asymptotically flat and maximal initial
  data with $N$ asymptotic ends. Denote by $M^i$, $J^i$ ($i=1,\ldots N$) the
  mass and angular momentum of the end $i$. Take an arbitrary end (say $1$),
  then the mass at this end satisfies the inequality
\begin{equation}\label{eq:49a}
 \mf(J^2,\ldots, J^N) \leq  M^1
\end{equation}
where $ \mf(J_2,\ldots, J_N)$ denotes the numerical value of the mass
functional $\mf$ evaluated at the corresponding harmonic map. 
\end{theorem}
This theorem reduces the proof of the inequality with multiples ends to compute
the value of the mass functional on the corresponding harmonic map. 

\cite{Khuri:2015yva} extend the result by \cite{Chrusciel:2007ak} in that electromagnetic 
charge is  
included and weaker fall off conditions are assumed on the extrinsic curvature. Also, the rigidity statement is proven.

Some numerical calculations have been made to get insight about the value $\mf(J_2,\ldots, J_N)$. \cite{Dain:2009qb} perform numerical calculations of the mass functional mentioned above (see section 
\ref{sec:proofMJ}) and find evidence for the validity of an intermediate inequality of the form $\mM\geq |J|$ where $J$ is the total angular
momentum of a system of two Kerr black holes with positive individual masses. Following this result, 
\cite{CabreraMunguia:2010uu} work on the Tomimatsu and Dietz--Hoenselaers solution describing two Kerr black holes, one of which has negative 
mass. They find that there is a rank in the parameters of the individual black holes such that the total mass is smaller than the total 
angular momentum, that is $M^2<|J|$ and therefore $\mM<|J|$, which opposes to Dain and Ortiz result.

\subsubsection{Non asymptotically flat manifolds}

Recently, global inequalities for asymptotically hyperboloidal initial data have started to be explored. 
\cite{Cha:2015nja} generalize a procedure used by \cite{Schoen81c} which consists in a 
deformation that
transforms an asymptotically hyperboloidal structure into an asymptotically flat one. By doing this, they are able to use the geometrical 
inequalities 
known to hold in the 
asymptotically flat case, to prove them in the hyperboloidal case. More precisely, they consider a smooth, simply connected, axially 
symmetric initial data satisfying the charged dominant energy condition $\mu_{M}\geq |J_{M}|$ and the matter condition $j_M^i\eta_i=0$. 
The data is assumed to have
two ends, one asymptotically hyperboloidal and the other either asymptotically flat or
asymptotically cylindrical. If certain system of equations, consisting of a Jang-like equation, and two extra equations on the deformed 
data, admits a smooth solution with prescribed asymptotics, then inequality \eqref{eq:42b} holds. Moreover, if equality is attained, then 
the initial data arise from an embedding into the extreme Kerr--Newman spacetime. This result effectively reduces the proof of the geometrical
inequality to proving existence of solutions to certain equations with appropriate fall off conditions.

\cite{Cha:2017gej} apply the same arguments to obtain an identical global inequality to \eqref{eq:42b} in the case that 
the initial 
data has two ends,
one AdS hyperbolic and the other either asymptotically AdS hyperbolic or
asymptotically cylindrical. The cosmological constant is not, however, explicitly included into the inequality. See section
\ref{sec:lambdam} for a conjectured global inequality including cosmological constant.

\subsubsection{The mass functional $\mM$}\label{sec:proofMJ}
We present in this section the main arguments behind the Mass-Angular momentum inequalities  \eqref{eq:42b} and \eqref{eq:49a}. 
They are heavily based on a certain mass functional $\mM$ and its relation with the energy of harmonic maps. 
For simplicity we assume electrovacuum and maximality to present the basic properties of $\mM$ that are crucial for proving \eqref{eq:42b}.
We also assume only two ends on the initial data.

The proof consists of two steps. The first one is to prove $m\geq\mM$ using the Hamiltonian equation and some energy conditions. The second step is to prove that extreme
Kerr--Newman black hole is a minimizer for $\mM$, that is $\mM\geq\mM|_{\mbox{\small{extr Kerr--Newman}}}$ using known results on harmonic maps
with prescribed boundary conditions. Let us see this in more detail.

\vspace{0.5cm}

\textbf{Step 1.} $m\geq\mM$

\vspace{0.5cm}

Consider an asymptotically flat, axially symmetric, maximal initial data set $(\Si, h_{ij}, K_{ij}, E^i, B^i)$. This means that
$K_{ij}h^{ij}=0$ and that there exists a rotational Killing vector field $\eta^i$ such that
\be
\pounds_\eta h=0, \quad \pounds_\eta K=0, \quad \pounds_\eta F=0, \quad \pounds_\eta {}^*F=0
\ee
where $\pounds$ is the Lie derivative.

Assuming $j^{EM}_i\eta^i=0$, the non trivial constraint equation reads
\be\label{lich1}
R=K_{ij}K^{ij}+16\pi(E_iE^i+B_iB^i)
\ee
and the Maxwell constraints without sources are
\be
dF=0,\quad d {}^*F=0
\ee
Due to axial symmetry, there exists a coordinate system such that the metric $h_{ij}$ can be written in the form
\be\label{eq:metricMJ}
h=e^{\sigma+2q}(d\rho^2+dz^2)+\rho^2e^{\sigma}(d\phi+\rho A_\rho+A_zdz)^2
\ee
where the functions $\sigma, q$ and $A_\rho, A_z$ depend only on $\rho,z$. See the article by \cite{Chrusciel:2007dd} where a careful
constructive proof of the existence of such coordinate system is done. The two ends correspond to the regions 
$r=\sqrt{\rho^2+z^2}\to \infty$, which is an asymptotically flat end, and $r\to0$, which is either asymptotically flat or cylindrical.

We follow \cite{Costa:2009hn}  to introduce the following 3-dimensional potentials. Let $\psi$ be the electric potential  (compare with the 4-dimensional potential $\psi$ 
presented in Sect.~\ref{sec:quasilocalq}) and $\chi$, 
the magnetic potential given by
\be
d\chi:=F(\bar\eta),\qquad d\psi:={}^*F(\bar\eta)
\ee
where $[F(\bar\eta)]_i=F_{ij}\eta^j$. We also introduce the potential $\omega$ as given by
\be
d\omega:=K(\bar\eta)\wedge\eta-\chi d\psi+\psi d\chi,
\ee
where, similarly, $[K(\bar\eta)]_i=K_{ij}\eta^j$.

These potentials have two properties that make them highly suited for the problem at hand. The first one is that $\psi, \chi$ and $\omega$ 
are constant on each connected component of the symmetry axis $\Gamma$. This, in particular, gives a close and simple relation
between these potentials and the physical quantities $Q_E$, $Q_B$, $J_\infty$:
\be
Q_{E\infty}=\frac{\psi_+-\psi_-}{2},\quad Q_{B\infty}=\frac{\chi_+-\chi_-}{2},\quad J_\infty=\frac{\omega_+-\omega_-}{8}
\ee
where the subindex $+$ and $-$ on a quantity $f$ indicate the constant values of the function on each connected component of $\Gamma$, 
namely: $f_+=f(\rho=0,z>0)$ and $f_-=f(\rho=0,z<0)$.

The second property is that they allow us to write the following important expressions
\be\label{boundK}
K_{ij}K^{ij}\geq 2\frac{e^{-3\sigma-2q}}{\rho^4}(\partial\omega+\chi\partial\psi-\psi\partial\chi)^2
\ee
and
\be\label{boundmu}
E_iE^i+B_iB^i\geq \frac{e^{-2(\sigma+q)}}{\rho^2}\left[(\partial\psi)^2+(\partial\chi)^2\right]
\ee
where in the left hand sides of \eqref{boundK} and \eqref{boundmu}, indices are moved with the metric $h_{ij}$ and the right hand sides  involve square gradients with respect to the flat 2-dimensional metric $d\rho^2+dz^2$.

Also, the curvature scalar $R$ of the metric $h_{ij}$ is bounded as
\be\label{scal1}
-\frac{1}{8}Re^{\sigma+2q}\geq\frac{1}{4}\Delta\sigma+\frac{1}{16}(\partial\sigma)^2+\frac{1}{4}\Delta_2q
\ee
where $\Delta$ is the flat Laplacian in 3-dimensions, $(\partial\sigma)^2=\frac{1}{\rho^2}(\partial_\rho\sigma)^2+(\partial_z\sigma)^2$
and  $\Delta_2:=\partial^2_\rho+\partial^2_z$.

Now we integrate the Hamiltonian equation \eqref{lich1} on $\mathbb R^3$ with the flat volume element $d^3x$ and use the bounds 
\eqref{boundK}, \eqref{boundmu} and \eqref{scal1} to obtain

 \be\label{lich2}
m\geq\frac{1}{32\pi}\int_{\mathbb R^3}(\partial\sigma)^2+ 4\frac{(\partial\omega+\chi\partial\psi-\psi\partial\chi)^2}{\eta^2}
+4\frac{(\partial\psi)^2+
(\partial\chi)^2}{\eta}\;d^3x
\ee
where we have used the expression of the ADM mass computed from \eqref{eq:massadm} for the particular metric \eqref{eq:metricMJ}
\be
m=-\frac{1}{8\pi}\int\Delta\sigma\; d^3x,
\ee
the explicit square norm of the axial Killing vector, $\eta=e^\sigma \rho^2$, and also the asymptotically flat  fall off conditions at 
infinity  and the regularity at the axis $q|_{\Gamma}=0$ to discard the term with $\Delta_2q$.

Defining the right hand side of \eqref{lich2} as the mass functional $\mM$
\be\label{defmM}
\mM:=\frac{1}{32\pi}\int_{\mathbb R^3}(\partial\sigma)^2+ 4\frac{(\partial\omega+\chi\partial\psi-\psi\partial\chi)^2}{\eta^2}
+4\frac{(\partial\psi)^2+
(\partial\chi)^2}{\eta}\;d^3x
\ee
we obtain the desired inequality $m\geq\mM$.

\vspace{0.5cm}

\textbf{Step 2.} $\mM\geq\mM|_{\mbox{\small{extr Kerr--Newman}}}$

\vspace{0.5cm}
We start by restricting the integral in the definition of $\mM$ to an open set $\Omega\subset\mathbb R^3$, and denoting the corresponding 
functional as $\mM_\Omega$. We also write it fully in terms of $\eta$ by taking into account that
\be
(\partial\sigma)^2=\left(\frac{\partial\eta}{\eta}-2\partial\ln\rho\right)^2=\left(\frac{\partial\eta}{\eta}\right)^2+
4\partial\ln\rho\partial\left(\ln\eta-\ln\rho\right).
\ee
We obtain
\begin{eqnarray}\label{defmMeta1}
\mM_\Sr=\frac{1}{32\pi}\int_{\Sr}\frac{(\partial\eta)^2}{\eta^2}+ 4\frac{(\partial\omega+\chi\partial\psi-\psi\partial\chi)^2}{\eta^2}
+4\frac{(\partial\psi)^2+
(\partial\chi)^2}{\eta}\;d^3x+\\
\frac{1}{8\pi}\int_{\Sr}\partial\ln\rho\partial\left(\ln\eta-\ln\rho\right)\;d^3x
\end{eqnarray}
We integrate by parts the last integral and use that $\ln\rho$ is harmonic ($\Delta\ln\rho=0$) to obtain

\begin{eqnarray}\nonumber
\mM_\Sr=\frac{1}{32\pi}\int_{\Sr}\frac{(\partial\eta)^2}{\eta^2}+ 4\frac{(\partial\omega+\chi\partial\psi-\psi\partial\chi)^2}{\eta^2}
+4\frac{(\partial\psi)^2+
(\partial\chi)^2}{\eta}\;d^3x+\\\label{defmMeta2}
\frac{1}{8\pi}\int_{\partial\Sr}\partial_s\ln\rho\left(\ln\eta-\ln\rho\right)\;d^3x
\end{eqnarray}
where $\partial_s$ denotes the derivative in the (outward) direction normal to $\partial\Sr$.

Here is where the connection with harmonic maps becomes evident. Recall that given the harmonic maps 
$(\eta,\omega,\chi\psi):\Omega\subset\mathbb R^3\setminus \Gamma\to\mathbb H^2_{\mathbb C}$, where $\Gamma$ is the 
symmetry axis and $\mathbb H^2_{\mathbb C}$ is the hyperbolic complex plane, the energy $\tilde\mM$ of such harmonic maps is defined by
\be\label{deftildemM}
\tilde\mM_\Omega:=\frac{1}{32\pi}\int_\Sr\frac{(\partial\eta)^2}{\eta^2}+ 
4\frac{(\partial\omega+\chi\partial\psi-\psi\partial\chi)^2}{\eta^2}
+4\frac{(\partial\psi)^2+
(\partial\chi)^2}{\eta}d^3x.
\ee
Therefore we have the relation
\be\label{mMtilemM}
\mM_\Omega=\tilde\mM_\Omega+B_{\partial\Omega}
\ee
where $B_{\partial\Omega}$ is the boundary term introduced  in \eqref{defmMeta2}.

We can apply the results of \cite{Hildebrandt77} (see also \citealp{Chrusciel:2007dd}) stating that when $\Omega$ is compact and does not contain 
the axis $\Gamma$, and the target manifold has negative curvature (as $\mathbb H^2_{\mathbb C}$ in our case), the minimizers of
$\tilde\mM_\Omega$ with Dirichlet boundary conditions exist, are unique, smooth and satisfy the Euler Lagrange equations.

Since the difference between $\mM_\Omega$ and the harmonic energy $\tilde\mM_\Omega$ is the boundary term $B_{\partial\Sr}$, 
the minimizer 
of $\tilde\mM_\Omega$ is the minimizer of $\mM_\Sr$ as well. This minimizer is the extreme Kerr--Newman solution, and we obtain
$\mM_\Omega\geq\mM_\Omega|_{\mbox{\small{ext Kerr--Newman}}}$. 
 
After a subtle limit procedure that allows to extend the inequality valid in $\Sr$ to all $\mathbb R^3$, one arrives at the
desired inequality $\mM\geq\mM|_{\mbox{\small{extr Kerr--Newman}}}$.

\subsection{Cosmological Constant}\label{sec:lambdam}

As we discuss in Sect.~\ref{subsec:MJ}, \cite{Cha:2017gej} study asymptotically AdS hyperbolic initial data and prove a 
global inequality where the cosmological constant does not appear explicitly. They conjecture, though, that an inequality of the form
\begin{eqnarray}
M\geq\frac{1}{3\sqrt{6}}\left[\sqrt{\left(1+\frac{J^2}{M^2}\right)^2+\frac{12 J^2}{M^2}}+2\left(1+\frac{J^2}{M^2}\right)\right]\times \\ 
\times\left[\sqrt{\left(1+\frac{J^2}{M^2}\right)^2+\frac{12 J^2}{M^2}}-\left(1+\frac{J^2}{M^2}\right)\right]^{1/2}
\end{eqnarray}
should hold, with equality in the extreme Kerr--Newman AdS black hole.

Related inequalities are given by \cite{Chrusciel:2006zs}, which is an extension of a previous work by \cite{Maerten:2006eua}.

\section{Quasilocal inequalities for black holes}\label{sec:QL}

The geometrical inequalities presented in this section relate purely quasilocal quantities 
defined on a closed 2-surface. By this we mean that no \textit{bulk} quantities defined on the 3-dimension region inside the surface are considered.

As is the case with global inequalities, here we divide the subject in three parts, the pure charge case, in Sect.~\ref{sec:AQ}, 
the inequalities 
involving angular momentum, in Sect.~\ref{sec:AJ} and the inequalities involving (explicitly) a cosmological constant in section 
\ref{sec:Lambda}. 
We do this for two reasons. First, the pure charge case problem does not need axial symmetry to be formulated, whereas it is needed in 
order to 
have a
well defined quasilocal angular momentum (see Sect.~\ref{sec:physical-quantities}). Also, even when the electric charge gives a 
flavor and a hint 
of what may happen with angular momentum, the techniques employed in the three treatments are usually different. 

Let us analyze the settings and the factors that produce these quasilocal inequalities.

\textit{Settings.} As opposed to the global inequalities, which can be proven for complete initial surfaces, quasilocal 
inequalities need a well identified surface representing the black hole, where things are computed. Even then, the problem can be formulated from two 
different perspectives: a Riemannian and a Lorentzian 
points of view. These two approaches have to do with the different surfaces used in the derivation of the inequality.
In the Riemannian setting, a minimal surface in an initial data set is studied. In the Lorentzian case, it is a MOTS in spacetime

\textit{Inequality producer.} In the quasilocal inequalities some form of a stability condition for the 2-surface considered is 
the main factor that produces the estimate. The Riemannian treatment
requires the positivity of the 
second variation of the area function. In the Lorentzian 
setting the inequalities
arise from stability of the MOTS.

\subsection{Area-Charge}\label{sec:AQ}

The relation between the area and the electromagnetic charges of a closed 2-surface was the first result obtained in the form of a 
quasilocal geometrical inequality.

It was first studied in a spacetime setting where trapping properties of 2-surfaces embedded in a Cauchy 
surface were assumed. Later, the inequality was proved purely from an initial data view point. And finally, from a purely Lorentzian one. 
We state a precise version of the 
result in the latter form below, and then discuss the different contributions, the hypotheses and 
conclusions. 

\subsubsection{Results}\label{sec:resultsAQ}

We extract the following theorem from the work of \cite{Dain:2011kb}. 
\begin{theorem}\label{thmAQ}
 Given an orientable closed marginally trapped surface $\Su$ satisfying the spacetime stably outermost condition, in a 
 spacetime which satisfies
Einstein equations, with non-negative cosmological constant $\Lambda$ and such
that the non-electromagnetic matter fields satisfy the dominant energy
condition, then the inequality
\be\label{eqAQ}
A\geq 4\pi Q^2
\ee
holds, where $A$ is the area of $\Su$ and $Q^2=Q_E^2 +Q_M^2$ is the total charge of $\Su$.
\end{theorem}
We  define the stability condition and address the other hypotheses below. Let us first review the different previous results 
on the subject.

\cite{Gannon76} obtains an inequality in the spirit of \eqref{eqAQ} when analyzing properties of  electrovacuum 
black hole spacetimes that are strongly 
future asymptotically predictable from a partial Cauchy surface $\Si$ regular near infinity. This means that 
$\Si=\cup_{i=1}^\infty \Sr_i$, with $\Sr_i\subset \Sr_{i+1}$ such that
$\partial \Sr_i:=\Su_i$ are future inner trapped surfaces (\textit{i.e.} $\theta_-<0$).
These 2-surfaces are expected to be found in most isolated gravitating systems after moving sufficiently far from the 
center of the system and are the alternative to stable MOTS and minimal surfaces used by later authors. 
In this setting Gannon proves that if the boundary of the black hole can be foliated by spacelike two-surfaces whose surface area 
is bounded above by $A_{max}$, then $A_{max}\geq 4\pi Q^2$, provided the electric charge $Q$ is not zero. Strictly speaking, Gannon's 
result is not quasilocal as it makes assumptions about all $\Si$. Nevertheless, we present the result here because of the great role it 
plays in  the subject as a model and inspiration to
subsequent work.


Later, \cite{Gibbons:1998zr}, using the positivity of the second variation of the area function, obtains the 
inequality \eqref{eqAQ} 
as a particular case of results in higher dimensions. For our purposes here, he considers maximal, electrovacuum initial data with a 
stable minimal surface $\Su$. Recall that given a maximal initial data $(\Si, h_{ij}, K_{ij}, \mu, j^i)$, and an embedded surface $\Su$ 
in $\Si$, we say it is a 
minimal surface if the mean curvature of $\Su$ vanishes and it is stable if the second variation of the area function is non negative, 
namely $\delta^2_{\alpha s}A\geq$ for all functions $\alpha$, where $s^i$ is a  normal vector to $\Su$ in $\Si$. The stability condition 
gives the desired inequality \eqref{eqAQ} between the area $A$ and the charge $Q$ of $\Su$. This result is a refinement of a 
previous work of \cite{Gibbons:1996af} in the spherically symmetric case. 


In the Lorentzian settings, \cite{Dain:2011kb} study the inequality between area and charge of a MOTS in a 
spacetime, satisfying certain stability property. Let us see it in some detail. 
\cite{Andersson:2005gq, andersson08} introduce a notion of stability for a closed marginally trapped 
surface $\Su$ which motivates the notion used in theorem \ref{thmAQ}. $\Su$ is said to be spacetime stably outermost if there exists
an outgoing vector $X^\mu=f_\ell\ell^\mu-f_k k^\mu$ with functions $f_\ell\geq0$ and $f_k>0$ such that the variation 
$\delta_{X}$ of $\theta_+$ with respect to $X^\mu$ satisfies $\delta_{X}\theta_+\geq0$, where $\delta$ is the variation operator 
associated with a deformation of $\Su$. 
Charged matter fields are included, although the non-electromagnetic matter fields must satisfy the
dominant energy condition.
Extensions of this inequality are
also proven for regions in the spacetime which are not necessarily black
hole boundaries, but ordinary objects (see section 
\ref{sec:objects} for more details). They prove theorem \ref{thmAQ}. They also prove  a similar inequality to \eqref{eqAQ} for an 
oriented surface screening an asymptotically flat end. A screening surface of an end is a closed 2-surface that encloses an open, connected 
region $\Sr\subset\Si$ which contains the mentioned end and no other.

As noted by \cite{Jaram12}, the area charge inequality relies on the algebraic properties of the electromagnetic energy 
momentum tensor. This observation provides a straightforward generalization of \eqref{eqAQ} to other  matter fields having similar 
algebraic properties. In particular, in \cite{Jaram12}, the inequality \eqref{eqAQ} is extended to include the Yang--Mills charges.

\subsubsection{Discussion}\label{sec:discussionAQ}
We wish now to make a few observations about this result.

As in the global case for the inequality between mass and charge (see the beginning of
Sect.~\ref{sec:global}), no axial symmetry is required to obtain \eqref{eqAQ} due to the area and charge being well defined 
quasilocal quantities.

On the other hand, as opposed to the global inequalities involving the ADM mass, \eqref{eqAQ} is a quasilocal relation, it refers to the properties of one 2-surface describing the horizon. This means that if 
one considers a spacetime containing many black holes, that inequality should hold for each one of them. The paradigmatic example 
of this case is the Majumdar--Papapetrou solution \citep{Majumdar47, Papapetrou45}  which consists of an arbitrary number of extreme Reissner--Nordstr\"om-like black holes,
all with charges of the same sign. This is a very special solution as each black hole saturates the bound \eqref{eqAQ}, i.e., $A_i=4\pi Q_i^2$ where 
$A_i$ and $Q_i$ are the individual areas and charges respectively of each black hole. Moreover, \eqref{eqAQ} is also saturated by the total values of horizon's 
area and charge of the solution.

Majumdar-Papapetrou solution is static, nevertheless, inequality \eqref{eqAQ} holds in completely dynamical scenarios as well, even in 
the presence of charged matter fields satisfying the dominant energy condition. This is in contrast with the Mass-Charge inequality 
presented in Sect.~\ref{sec:MQ}, where a strong (and in a sense, unnatural) local condition is imposed on matter fields (see the discussion after
equation \eqref{eq:37}). 

Note  that there is no rigidity statement saying that if equality is attained in \eqref{eqAQ}, then the solution must be the near 
horizon geometry of Reissner Nordstr\"om. 

In fact, the key ingredient used to prove theorem \ref{thmAQ} is the notion of stability for the 2-surface. This condition plays 
an analogous role as the non-negativity condition on the second variation of area function in Riemannian settings.

\subsection{Area-Angular Momentum-Charge}
\label{sec:AJ}

At the beginnings of 2007, two results relating horizon area and angular momentum of black holes were given, one for stationary black
holes by Ansorg and Pfister and the other for isolated horizons by Booth and Fairhurst (see the Living Review by  
\citealp{Ashtekar04} for definitions and general results on isolated horizons). These two works have motivated the more general  study of 
quasilocal inequalities that explicitly include the angular momentum of a given surface in a dynamical system.

The different settings where the desired inequality have been proven are stationary spacetimes,  maximal 
initial data and finally trapped surfaces.

\subsubsection{Results}\label{sec:resultsAJ}
We present here one of the results, taken from the article by \cite{Clement:2012vb} that is valid for dynamical as well as stationary black holes represented by an appropriate
closed 2-surface.

\begin{theorem}\label{t:AJQ}
 Let $\Su$ be either
\begin{enumerate}
 \item a smooth spacetime stably outermost axisymmetric marginally outer trapped surface (MOTS) embedded in a
spacetime, satisfying the dominant energy condition, or
\item  a smooth stable axisymmetric minimal surface in a maximal data set, with non-negative
scalar curvature,
\end{enumerate}
with a non-negative cosmological
constant $\Lambda$, angular momentum $J$, charges $Q_E$ and $Q_B$ and area $A$. Then,
\be\label{AP2}
A\geq\sqrt{(8\pi J)^2+(4\pi Q^2)^2}
\ee
with $Q^2=Q_E^2+Q_B^2$. 

Moreover, the equality in \eqref{AP2} is achieved if and only if the surface is the extreme Kerr--Newman
sphere.
\end{theorem}

We review the many results that led to this theorem and discuss the theorem afterwards, in Sect.~\ref{sec:discussionAJ}.

The first, known to us, work on a geometrical inequality in the spirit of \eqref{AP2} is due to \cite{Ansorg:2007fh}. 
They treat stationary and  degenerate black holes and find that they must satisfy the equality in \eqref{AP2}. 
More precisely, the equality in \eqref{AP2} holds for every element in a parametric sequence of axially and equatorially symmetric, 
stationary systems 
consisting of a degenerate black hole surrounded by matter such that the limit system is the Kerr--Newman black hole. 
Moreover, the authors conjecture that for axially and equatorially symmetric, stationary black holes surrounded by matter, the 
inequality \eqref{AP2} should hold, with equality at the degenerate case.

A few months later, \cite{Booth:2007wu} argue that the allowed values for the angular momentum of an isolated 
horizon $\Su$ should be determined from the intrinsic horizon geometry. They find
\be\label{BF}
2\sqrt{e\gamma}A\geq8\pi|J| 
\ee
where $e$ is the surface integral of the evolution equation for the inward expansion at the horizon and $\gamma:=\pi A^{-2}\int_S\eta$,
where $\eta=\eta^i\eta_i$ is the square norm of the rotation vector field. Moreover, $e\leq1$ and  $e=1$ at an extremal horizon and 
$\gamma<1/4$ for axially symmetric horizons whose cross sections can be embedded in Euclidean space. In that case one obtains the strict inequality
\be
A>8\pi|J|.
\ee
However, the authors argue that otherwise, $\gamma$ can become arbitrarily large making the bound \eqref{BF} to lose its meaning. 

The Ansorg and Pfister conjecture  is finally proven in vacuum by \cite{hennig08, Hennig:2008zy}, and \cite{Ansorg:2010ru}. They 
show that every axially symmetric and stationary black hole with surrounding matter satisfies \eqref{AP2}
and equality holds if the black hole is extremal. The proof consists in showing that if $A\leq\sqrt{(8\pi J)^2+(4\pi Q^2)^2}$, then the 
black hole can not be subextremal in the sense of \cite{Booth:2007wu}. Recall that a black hole is subextremal if 
there exist trapped surfaces in every small interior vicinity of the event horizon. Einstein equations near the horizon are then considered
and a variational problem is formulated and solved. As we discuss in Sect.~\ref{sec:discussionAJ} this stationary variational problem is 
closely related to the variational problem arising in the dynamical regime.

\cite{dain10d} conjectures the validity of the vacuum case of \eqref{AP2} for the connected component of the apparent horizon 
in a dynamical scenario. The proof was given later, as in the pure charge case, from two perspectives, one Riemannian and one Lorentzian.

With a Riemannian approach, \cite{Acena:2010ws} prove that extreme Kerr initial data is the 
global minimum for certain mass functional $\mathbb M$ related to the second variation of the area functional and analogous to 
the mass functional $\mM$
presented in Sect.~\ref{sec:proofMJ}. 
They prove the inequality \eqref{AP2} with $Q=0$ for vacuum axially symmetric initial data containing a minimal surface and 
such that the metric 
$h_{ij}$
satisfies certain 
technical conditions. 
However, this class considered includes many known black hole initial data.
They extend the validity of the inequality to include a non-negative cosmological 
constant, not appearing explicitly into the inequality though. However, this generalization is relevant because 
there exists a counter-example of the inequality (\ref{AP2}) with $Q=0$ for the case of negative cosmological 
constant, as it was pointed out in \cite{Booth:2007wu}.

In \cite{Clement:2011kz}, the author relaxes some of the restriction on the type of surfaces studied and  
\cite{Dain:2011pi} prove 
the inequality \eqref{AP2} with $Q=0$, in vacuum, replacing the technical conditions of \cite{Acena:2010ws}  by the stability condition on the 
minimal surface (see Sect.~\ref{sec:resultsAQ} for definition of stable minimal surface).

With a Lorentzian treatment, \cite{Jaramillo:2011pg} prove \eqref{AP2} for an axially symmetric 
closed marginally
trapped surface $\Su$ satisfying the spacetime stably outermost condition (see Sect.~\ref{sec:resultsAQ}), in a spacetime with
non-negative cosmological constant and matter fields satisfying the dominant energy condition. The rigidity statement with the extreme Kerr 
sphere, instead of the extreme Kerr--Newman sphere is also proven for 
the $Q=0$ case.

The inclusion of electromagnetic charges is done by \cite{Clement:2011np} and by \cite{Clement:2012vb}. Through the introduction of appropriate electromagnetic and angular momentum potentials, they 
prove theorem \ref{t:AJQ} in two different ways and show the connection with the variational problem for the stationary case (see the 
remarks below). Also a  
relation between the two mass functionals $\mM$ introduced in Sect.~\ref{sec:proofMJ}, equation \eqref{defmM} and $\mathbb M$ is 
pointed, which suggests a relation between the global inequality \eqref{eq:42b} and the quasilocal inequality \eqref{AP2}. We explore 
this in Sect.~\ref{sec:proofAJ} with more detail.

\subsubsection{Discussion} \label{sec:discussionAJ}

We wish to make a few remarks about the hypotheses and statements of theorem \ref{t:AJQ}. 

Theorem \ref{t:AJQ} calls for stable closed surfaces. As we mention at the beginning of Sect.~\ref{sec:QL}, it is this
stability property what drives the inequalities. Both concepts of stability appeared already in 
problem with no angular momentum, see Sect.~\ref{sec:AQ} for definitions. The only extra assumptions we make in this result  is that the functions $\alpha$ and $f_\ell, f_k$ entering the stability criteria for minimal surfaces and MOTSs respectively
must be axially symmetric.  
In Sect.~\ref{sec:proofAJ} the connection between these two stability conditions is revised. 
The interesting point made in \cite{Clement:2012vb} is that the corresponding stability assumptions both for minimal surfaces and MOTS lead
to exactly the same integral condition.

The theorem admits a non negative cosmological constant, but it does not enter explicitly into the inequality. The treatment of such problem
needs different techniques and is reviewed in Sect.~\ref{sec:Lambda}. The relevant property of the cosmological constant in this theorem 
is that its positivity allows one to disregard it altogether from the Einstein or constraint equations. Clearly, the same can not be made 
for negative $\Lambda$ and it is still a very important open problem. 

Non electromagnetic matter fields are admitted in the hypotheses of theorem \ref{t:AJQ} and they are not required to satisfy the dominant 
energy condition. Indeed, the complete energy momentum tensor \textit{i.e.} $T_{\mu\nu}^M+T_{\mu\nu}^{EM}$ must. In particular there can be matter surrounding and 
crossing the surface $\Su$. This in particular extends the results in
\cite{Ansorg:2010ru} and \cite{Hennig:2008zy}.

A major difference between theorems \ref{thmAQ} and \ref{t:AJQ} is the rigidity statement. The horizon in extreme Reissner--Nordstr\"om 
clearly saturates inequality \eqref{eqAQ}, but so far it is not proven that it is the only horizon that does. On the other hand, the horizon in extreme Kerr--Newman 
is the unique solution that satisfies the equality in \eqref{AP2}. In the latter case, it is the connection between a certain mass functional
$\mathbb M$ and the energy of harmonic maps and the uniqueness of minimizers of that energy what ultimately gives uniqueness in theorem 
\ref{t:AJQ}. See Sect.~\ref{sec:proofAJ} for details. 

The extreme Kerr--Newman sphere mentioned in theorem \ref{t:AJQ} has a precise meaning in terms of intrinsic and extrinsic quantities defined 
on the surface $\Su$ \citep{Clement:2012vb}. Basically the surface has the geometry of a horizon section in the extremal Kerr--Newman 
black hole. In section 
\ref{sec:proofAJ} we give proper definitions, here, however, we want to emphasize that the rigidity statement refers to the extreme 
Kerr--Newman horizon, not the entire initial data. Interestingly, the fact that the equality in \eqref{AP2} is only attained by 
the extreme Kerr--Newman sphere has been known since the work of \cite{Hajicek:1974oua}, 
\cite{Lewandowski:2002ua}, and  
\cite{Kunduri:2008rs}, \cite{lrr-2013-8} on isolated horizons and near horizon geometries of extreme 
black holes. See also the more recent results of \cite{Reiris:2012en} and \cite{Chrusciel:2017vie}).

Finally we want to mention the relation between the variational problem used to prove theorem \ref{t:AJQ} and the one used in the
stationary case in \cite{Ansorg:2010ru}.  The argument in \cite{Ansorg:2010ru} to prove the strict 
inequality \eqref{AP2} is based on the implication
\be\label{statarg}
\mbox{subextremal horizon}\quad \Rightarrow \quad A>\sqrt{(8\pi J)^2+(4\pi Q^2)^2}.
\ee
The counterreciprocal of \eqref{statarg} is written as a variational problem for an action functional on a Killing horizon section. As it 
is shown in \cite{Clement:2011np}, this action and variational problem are identical to the corresponding mass functional $\mathbb M$ and the variational
problem formulated for a stable MOTS.  This connection is particularly remarkable. We mentioned already a link between 
the variational problems for stable minimal surfaces and 
stable MOTS. This is essentially  a manifestation of the close relation between the two different characterizations of black holes. However,
the great similarities with the stationary case are not at all a priori obvious, especially considering that the treatment in \cite{Ansorg:2010ru} makes use 
of the particular form of the 4-dimensional stationary, axially symmetric metric, whereas the arguments in the proof of theorem 
\ref{t:AJQ} refer solely to the stable surface $\Su$.

\subsubsection{Area products}
We want to mention a close relation valid for axially 
symmetric, charged, rotating and stationary 
black holes  with surrounding matter. It not only involves the event horizon area, $A$,
but also the Cauchy horizon area $A_{\mbox{\small{Cauchy}}}$. It reads
\be\label{areaprod}
(8\pi)^2\left(J^2+\frac{Q^4}{4}\right)=AA_{\mbox{\small{Cauchy}}}
\ee
The remarkable observation is that the area product does not depend on the total mass, equation \eqref{areaprod} is quasilocal. 
Is is a consequence of  the fact that there can not be matter between the event and Cauchy horizon due to 
stationarity.  
Equation \eqref{areaprod} has been proven by \cite{Ansorg:2008bv,
  Hennig:2009aa, Ansorg:2009yi} and by
\cite{Ansorg:2010ru}. See also Visser's work \citep{Visser:2012wu} on the validity of such mass independent, area-related functions . It also received a huge interest in string and other theories, see the work  by
\cite{Cvetic:2010mn} and the review by \cite{Compere:2012jk} on the Kerr/CFT correspondence.

\subsubsection{Shape of black holes}

A closely related quasilocal inequality for black holes is obtained by \cite{Reiris:2013jaa}. They link black hole 
shape parameters with angular momentum. More precisely, for a rotating, axially symmetric spacetime stably outermost horizon, the 
length $\mathcal C_e$ of the greatest axially symmetric circle 
and the length of the meridian $L$ satisfy
\be
\frac{16\pi^2|J|}{\sqrt{4\pi A}}\leq \mathcal C_e\leq \sqrt{4\pi A}
\ee
\be
4|J|\leq\frac{A}{2\pi}\leq L^2
\ee
and
\be
\frac{A}{L^2}\leq \frac{\mathcal C_e}{L}\leq 2\sqrt{2}\pi.
\ee
There are three effects that show up in these results. The most expected one is a thickening of the bulk of the horizon due to rotation. 
They also show that rotation stabilizes the horizon's shape  in that the area and angular momentum control completely its local shape. Finally,
at high angular momentum, the geometry of the horizon goes to that of extreme Kerr horizon, even in non vacuum.

\subsubsection{The mass functional $\mathbb M$}\label{sec:proofAJ}
Analogous to the global case, the proof of the quasilocal inequality \eqref{AP2} is based on some remarkable properties of a quasilocal 
mass functional $\mathbb M$ and consists of two intermediate inequalities for $\mathbb M$. 
Again, for simplicity, we assume electrovacuum.

The proof consists of two steps. The first one is to prove a lower bound on the area of the 2-surface $\Su$ in terms of a mass functional 
$\mathbb M$. One starts with the appropriate stability condition (for either type of surface, minimal or marginally outer trapped) and 
proves $A\geq 4\pi e^{\frac{\mathbb M-8}{8}}$. The second step is to prove that extreme
Kerr--Newman horizon is the unique minimizer for $\mathbb M$, that is $\mathbb M\geq\mathbb M|_{\mbox{\small{extr Kerr--Newman}}}$. 
We follow  \cite{Clement:2012vb}.

\vspace{0.5cm}

\textbf{Step 1.} $A\geq 4\pi e^{\frac{\mathbb M-8}{8}}$

\vspace{0.5cm}

For an axially symmetric, stable, minimal surface $\Su$, the stability  condition $\delta^2_{\alpha s}A\geq0$ is written in an integral 
form as
\be\label{minstab}
\int_{\Su}|D\alpha|^2+\frac{{}^2R}{2}\alpha^2 \; \ds\geq\int_{\Su}\frac{1}{2}({}^3R+|\Theta|^2)\alpha^2\; \ds
\ee
for arbitrary, axially symmetric functions $\alpha$, and where ${}^2R$ and ${}^3R$ are the scalar curvature of the metrics on $\Su$ and on $\Si$ respectively. $\Theta$ is the traceless part 
of the extrinsic curvature of $\Su$. The norms and surface element $\ds$ are computed with respect to the intrinsic metric on $\Su$.

On the other hand, for an axially symmetric, spacetime stably outermost MOTS $\Su$, the stability  condition $\delta_{X}\theta_+\geq0$ 
with $X^\mu=\alpha\ell^\mu+\Psi k^\mu$ where $\alpha>0$ and $\Phi\geq0$ are axially symmetric arbitrary functions, can be written (after
using Einstein equations and disregarding terms with the appropriate sign)
\be\label{motsstab}
\int_{\Su}|D\alpha|^2+\frac{{}^2R}{2}\alpha^2 \; \ds\geq\int_{\Su}+|\Upsilon^{(\eta)}|^2+E_\perp^2+B_\perp^2\; \ds
\ee
where $\Upsilon^{(\eta)}$ is the projection on the axial Killing vector $\eta$ of the normal fundamental form of $\Su$ and 
$E_\perp:=\ell^\mu k^\nu F_{\mu\nu}$, $B_\perp:=\ell^\mu k^\nu {}^*F_{\mu\nu}$ are the electromagnetic fluxes across the surface $\Su$.

The two inequalities \eqref{minstab} and \eqref{motsstab} become identical when the Hamiltonian constraint \eqref{lich1} is inserted in 
\eqref{minstab} and the relation $\Upsilon^{(\eta)}_\mu\eta^\mu=-K_{\mu\nu}\eta^\mu s^\nu$ is considered. Here $s^\mu$ is a spacelike normal to $\Su$.

One starts by writing the metric on the surface as (see \citealp{Dain:2011pi} where such coordinate system is constructed)
\be
ds^2=e^{2c-\sigma}d\theta^2 +e^{\sigma}\sin^2\theta d\varphi^2
\ee
where $\sigma$ is a function depending only on $\theta$ and  $c$ is a constant related to the area of $\Su$ and $\sigma$ by 
$A=4\pi e^c$ and 
\be\label{sigmapot}
\sigma|_{\theta=0,\pi}=c.
\ee

In these coordinates, the axial Killing vector field on $\Su$ is $\eta^i=\partial_\varphi^i$ and its square norm is given by
$\eta=e^\sigma\sin^2\theta$. The component $\Upsilon^{(\eta)}$ of the normal form can be written in terms of a function $\tilde\omega$ as
\be
\Upsilon^{(\eta)}_\theta=0,\quad \Upsilon^{(\eta)}_\varphi=-e^{\sigma-c}\sin\theta \tilde\omega'.
\ee
where a prime denotes derivative with respect to $\theta$.

As in the proof of the global inequality, suitable potentials $\psi, \chi, \omega$ for the electromagnetic fields and rotation 
are introduced via the equations
\be\label{psichi}
\psi'=-E_\perp e^c\sin\theta,\qquad \chi'=-B_\perp e^c\sin\theta
\ee
\be\label{omega}
\omega'=2\eta\tilde\omega'-2\chi\psi'+2\psi\chi'.
\ee
Note that we use the same letters to denote these 2-dimensional potentials and the 3-dimensional potentials introduced in section 
Sect.~\ref{sec:proofMJ}. As the latter, the potentials defined by \eqref{psichi}, \eqref{omega} have the important property
\be\label{boundaryvalues}
Q_E=\frac{\psi(\pi)-\psi(0)}{2},\quad, Q_B=\frac{\chi(\pi)-\chi(0)}{2},\quad J=\frac{\omega(\phi)-\omega(0)}{8},
\ee
where the charges and angular momentum refer to the surface $\Su.$

Writing the stability condition in terms of these potentials, and setting the arbitrary function $\alpha$  to be
\be
\alpha=e^{c-\sigma/2}
\ee
(see \citealp{Clement:2012vb} for a discussion about this choice) one finds 
\be
\frac{A}{4\pi}\geq e^{\frac{\mathbb M-8}{8}}
\ee
where the mass functional $\mathbb M$ is defined by
\be\label{defmbm}
\mathbb M:=\frac{1}{2\pi}\int\left[4\sigma+(\sigma')^2+\frac{(\omega'+2\chi\psi'-2\psi'\chi')^2}{\eta^2}+
4\frac{(\psi')^2+(\chi')^2}{\eta}\right]\ds_0
\ee
where the norms and surface element are computed with respect to the round metric on the unit sphere $d\theta^2+\sin^2\theta d\varphi^2$.
The great resemblance between $\mathbb M$ and the mass functional $\mM$ defined in \eqref{defmM} is discussed in Sect.~\ref{relation}.

One of the most remarkable property of the functional $\mathbb M$ is that the boundary conditions for the functions it
depends on are the angular momentum  and charges \eqref{boundaryvalues} and the area \eqref{sigmapot}. This is especially relevant for the 
formulation and solution of the variational problem in the next step.

\vspace{0.5cm}

\textbf{Step 2.} $\mathbb M\geq \mathbb M|_{\mbox{\small{extr Kerr Newman}}}$

\vspace{0.5cm}

The second step is the 
resolution of a variational principle for the functional $\mathbb M$ giving the global minimum in terms of the angular 
momentum and charges:
\begin{equation}\label{Mine}
e^{\frac{{\mathbb{M}}-8}{4}}\geq\ 4J^{2} + Q^{4},
\end{equation}
with $Q^2=Q_{E}^2 + Q_{B}^2$.

The proof of the inequality \eqref{Mine} can be approached in several ways, as presented in \cite{Clement:2011np} and \cite{Clement:2012vb}. We already 
commented the reduction to the variational problem in stationary settings. Here we mention the other two arguments.

One of them follows the lines of Step 2 in Sect.~\ref{sec:proofMJ}. Here, the energy $\tilde{\mathbb M}$ of harmonic maps 
$(\eta,\omega,\chi,\psi):U\subset S^2\setminus\{\theta=0,\pi\}\to \mathbb H^2_{\mathbb C}$ is considered. It reads
\be\label{deftildemb}
\tilde{\mathbb M}_U:=\frac{1}{2\pi}\int\left[\frac{(\eta')^2}{\eta^2}+\frac{(\omega'+2\chi\psi'-2\psi'\chi')^2}{\eta^2}+
4\frac{(\psi')^2+(\chi')^2}{\eta}\right]\ds_0.
\ee
Then, restricting the integral in the definition of $\mathbb M$ to the region $U$ and denoting it by $\mathbb M_U$, we obtain the relation
\be
\tilde{\mathbb M}_U=\mathbb M_U+4\int_U\ln\sin\theta\, \ds+\oint_{\partial U}(4\sigma+\ln\sin\theta)\partial_\nu\ln\sin\theta\, dl
\ee
where $\partial_\nu$ is the derivative in the direction of the  exterior unit vector normal $\nu$ to $\partial U$ and $dl$ is the line 
element in $\partial U$. We see that the difference 
between $\mathbb M_U$ and $\tilde{\mathbb M}_U$ is a constant plus a boundary term, which implies that both functionals have the same 
Euler--Lagrange equations. The result of \cite{Hildebrandt77} is again used to give existence of a
unique minimizer for $\tilde {\mathbb M}_U$. That minimizer is the extreme Kerr--Newman sphere, defined as the set 
$(\sigma_0,\omega_0, \chi_0,\psi_0)$ that can be obtained by computing the geometry on a horizon section of the extreme Kerr--Newman solution.

Finally a very subtle limit procedure must be performed to arrive at the desired inequality $\mathbb M\geq\mathbb M|_{\mbox{\small{extr Kerr Newman}}}$.

It is worth mentioning that the previous variational problem can be solved without assuming axial symmetry on the functions $\sigma,\omega, \psi,\chi$.

The second approach is restricted to axial symmetry and hence in the minimization problem for 
$\mathbb M$, the Euler-Lagrange equations reduce to a system of ordinary differential equations. When solving these equations, 
the boundary conditions $J$, 
$Q_{E}$ and $Q_{ M}$ determine uniquely the boundary conditions for the 
remaining potential $\sigma$.  This is the key fact under the sharpness of inequality \eqref{AP2}. A constructive explicit proof of 
existence and uniqueness for the minimizer of $\mathbb M$ is given in \cite{Clement:2012vb} with prescribed values of 
$J, Q_{ E}, Q_{ M}$ and without any reference to the boundary values of $\sigma$. This is different to what one does 
in the first approach discussed above, where the boundary values of $\sigma$ are prescribed from the relation 
$A=4\pi e^{\sigma}|_{\theta=0}$ valid for the particular coordinate system employed.

\subsection{Relation between $\mM$ and $\mathbb M$}\label{relation}

It is remarkable that both, the global \eqref{eq:42b} and the quasilocal \eqref{AP2}
inequalities involving angular momentum are derived from mass functionals $\mM$ and $\mathbb M$ respectively, and that these functionals are
minimized by some form of the extreme Kerr--Newman solution. 
In \cite{Clement:2012vb}, a connection between these two functionals is presented, which in turn, gives a connection between the 
two inequalities.

More precisely, it is shown there that the inequality $m\geq\mM\geq\mM_0$ implies that the
extreme Kerr--Newman horizon is a local minimum of the mass functional $\mathbb M$, which suggests that the global inequality \eqref{eq:42b} implies
the quasilocal inequality \eqref{AP2}:
\be
M^2\geq  \frac{Q_\infty^2+\sqrt{Q_\infty^4+4J_\infty^2}}{2} \quad\Rightarrow \quad A\geq\sqrt{(8\pi J)^2+(4\pi Q^2)^2}
\ee

On the other hand, Penrose inequality together with the quasilocal inequality \eqref{AP2} give
\be
\left[M^2\geq\frac{A}{16\pi}\right]\quad +\quad\left[ A\geq\sqrt{(8\pi J)^2+(4\pi Q^2)^2}\right]\quad \Rightarrow \quad M^2\geq\frac{A}{16\pi}\geq\frac{4J^2+Q^4}{4}
\ee
for stable MOTS, which is a weaker version of the global inequality \eqref{eq:42b}.

Whether there exist a deeper connection and a full implication of the form
\be
M^2\geq \frac{Q^2+\sqrt{Q^2+4J^2}}{2} \quad \Longleftrightarrow\quad A^2\geq(8\pi J)^2+(4\pi Q^2)^2
\ee
is far from settled. See also the discussion given in Sect.~\ref{sec:phys-interpr-stat} about this issue in the context of stationary 
black holes.

\subsection{Cosmological Constant}
\label{sec:Lambda}
The results of Sects.~\ref{sec:AQ} and \ref{sec:AJ} admit non negative cosmological constants, but do not include them 
explicitly into the inequalities. These results are presented in this section as, in general, they require different techniques.

By analyzing explicit solutions and collapsing black holes, \cite{Hayward:1993tt} and \cite{Shiromizu:1993mt}, prove that a positive cosmological constant sets restrictions 
on how large a black hole can be (see also \citealp{Maeda:1997fh}). They study black hole spacetimes with positive cosmological constant $\Lambda$, that satisfy the dominant
energy condition, and find that the area of the black hole horizon, as described by an outer marginal surface, is bounded as $A\leq4\pi/\Lambda$. 
The same inequality holds for the area of a connected section of the event horizon in the case of strongly future asymptotically predictable, 
asymptotically de Sitter spacetime. The inequality is saturated for the extreme Schwarzschild-deSitter horizon.

For negative $\Lambda$, \cite{Gibbons:1998zr} (in the time symmetric settings) and \cite{Woolgar:1999yi} (in the 
non time symmetric case) find the bound $A>4\pi(g-1)/|\Lambda|$, 
for the area of an outermost MOTS of genus $g>1$.

These inequalities show the important role that the cosmological constant plays in determining the size of a black hole.
Note in particular, that the positive and negative cosmological constants bound the area in opposite directions. Namely a de Sitter-like 
black hole can not be too large and an anti de Sitter-like black hole can not be too small. This has interesting implications for studying 
possible colliding scenarios. 

\cite{Gibbons:1998zr} also considers the combined effect of a cosmological constant and matter fields satisfying the dominant 
energy condition. He finds that the area of a stable minimal surface $\Su$ in a time symmetric 3-surface is bounded as 
$4\pi(1-g)-\Lambda A-\int_S8\pi T_{00}>0$, where $T_{00}$ is the energy density of the matter fields on the 3-surface. 

\cite{Simon:2011zf} arrives at the same inequality for stable MOTS in a spacetime satisfying the dominant energy condition. Interestingly,
when Maxwell fields are explicitly taken into account, he is able to write the inequalities as

\be\label{Simon1}
2\pi(1-\sqrt{1-4\Lambda Q^2})\leq \Lambda A\leq 2\pi(1+\sqrt{1-4\Lambda Q^2})\qquad \Lambda>0,
\ee
\be
\label{Simon2}
2\pi[g-1+\sqrt{(g-1)^2-4\Lambda Q^2}]\leq -\Lambda A\qquad \Lambda<0.
\ee
Note that \cite{Simon:2011zf} maintains the (non-negative) principal eigenvalue of the stability operator in his inequality. We omit it here for simplicity.
Inequalities \eqref{Simon1} show that for positive cosmological constant one obtains both an upper and a lower bound to the area. This in 
essence is a manifestation of a competition of two effects. On one hand, the charges forbid the black hole to become too small (due to 
electric repulsion). On the other hand, the positive cosmological constant forbids it to become too large (acting as a cosmological attraction). 
Inequality \eqref{Simon2} shows that when $\Lambda$ is negative, both effects, the 'cosmological' and the electric repulsion combine 
to give a lower bound to the area. 

The inequalities \eqref{Simon1}-\eqref{Simon2} are saturated in spherical symmetry by the Reissner--Nordstr\"om--de~Sitter black holes if and 
only if the surface gravity vanishes. Simon also discusses the time evolution of 
MOTSs and the application of his inequalities as restrictions on the merging. He 
deduces an interesting Corollary, non trivial only when $\Lambda\neq0$, which gives lower and upper bounds on the
quotient between the \textit{initial} and \textit{final} areas of MOTSs (homologous MOTSs). This result is 
applied to the situation where only a single MOTS is \textit{initially} present, and to the problem of merging of MOTSs.

As has been previously pointed out, the inclusion of angular momentum into geometrical inequalities requires different techniques. 
The extension of \eqref{Simon1} to rotating black holes was done by  
\cite{Clement:2015fqa}. They consider an axially symmetric, stable MOTS, with $\Lambda>0$ and matter satisfying the DEC, and
find that the allowed values of angular momentum $J$ are given by
\be\label{GabachL1}
|J|\leq \frac{A}{8\pi}\sqrt{\left(1-\frac{\Lambda A}{4\pi}\right)\left(1-\frac{\Lambda A}{12\pi}\right)}
\ee
where $A$ is the area of the MOTS. The inequality \eqref{GabachL1} is saturated by the extreme Kerr-deSitter horizons.
One can  read from \eqref{GabachL1} that the presence of a positive cosmological constant sets stronger limits to the allowed values of 
the angular momentum. Namely, a cosmological horizon must rotate more slowly than the non cosmological one. This observation agrees with the 
intuitive idea mentioned above, that the cosmological constant has an attractive effect. Hence, if the area is fixed, then the rotation 
must be slowed down. Another way of looking at \eqref{GabachL1} is to consider the right hand side of \eqref{GabachL1} as an effective
area $A_{eff}$, with $A_{eff}\leq A$ due to being $\Lambda>0$. In this notation, \eqref{GabachL1} reads $|J|\leq A_{eff}/8\pi$.
It is worth remarking that the proof of inequality \eqref{GabachL1} follows the lines of the proof the Area-Angular momentum inequality (see section 
\ref{sec:proofAJ}), in that the stability condition is used to obtain a lower bound on the area in terms of a mass 
functional 
$\mathbb M^\Lambda$ defined by 
\be
\mathbb M^\Lambda(\sigma,\omega, A,a):=\frac{1}{2\pi}\int\left[\sigma'^2+\frac{\omega'^2}{\eta^2}+4\sigma\frac{1+\Lambda a^2\cos^2\theta}{\zeta}+
4\left(\frac{A}{4\pi}\right)^2\Lambda e^{-\sigma}\right]\zeta\; d^3x
\ee
where
\be
\zeta:=1+\frac{\Lambda a^2\cos^2\theta}{3}
\ee
and we have explicitly written the elements that $\mathbb M^\Lambda$ depends on, because they make the variational principle much harder 
than when one bounds away the cosmological constant (as was shown in Sect.~\ref{sec:proofAJ}). The first difficulty that arises 
when one keeps the term containing $\Lambda$ (to ensure that it will come up in the final inequality) is that the mass functional also 
depends explicitly on $A$

The second difficulty is proving existence and uniqueness of a minimizer for $\mathbb M^\Lambda$, as there is no direct relation between 
$\mathbb M^\Lambda$ and energy of harmonic maps. These obstacles are overcome as follows. The first one is dealt with by a scaling argument where $A$ and
$J$ are frozen to the extreme
Kerr-deSitter values and the dynamical variables in $\mathbb M^\Lambda$ change appropriately. For the second one, 
it is proven that every critical point of $\mathbb M^\Lambda$ is a local minimum and then the mountain pass theorem
is used to obtain the global existence.
In the presence of Maxwell fields, the inequality 
\be\label{GabachL2}
J^2\leq \frac{A^2}{64\pi^2}\left[\left(1-\frac{\Lambda A}{4\pi}\right)\left(1-\frac{\Lambda A}{12\pi}-\frac{2\Lambda Q^2}{3}\right)\right]-\frac{Q^4}{4}
\ee
is conjectured to hold in \cite{Clement:2015fqa} under the same hypotheses.

Inequality \eqref{GabachL2} was proven by \cite{Bryden:2016frb} following the same ideas as in \cite{Clement:2015fqa},
but 
simplifying the resolution of finding the minimizer of $\mathbb M^\Lambda$. The argument is based on the result of \cite{Schoen:2012nh},
which states that $\mathbb M^\Lambda$ is convex along geodesic deformations within $\mathbb H^2_{\mathbb C}$.

The case of negative cosmological constant is considerably more complicated as $\Lambda$ appears with the wrong sign in the mass functional $\mM$.
In a different context, \cite{Kunduri:2008rs} and \cite{lrr-2013-8} prove that the near horizon geometry of 
axisymmetric and stationary black holes is the one of the extremal
Kerr--Newmann-anti deSitter horizon and therefore they saturate \eqref{GabachL2} (see \citealp{Hennig:2014wqa} for an explicit 
expression and discussion). 

\subsection{An application: Non-existence of two black holes in equilibrium}
A very interesting application of the area-angular momentum inequality \eqref{AP2}, with $Q=0$, is the result by  
\cite{Neugebauer:2009su, Hennig:2011fp, Neugebauer:2011qb}, and \cite{Neugebauer:2013ee} where they prove that there does 
not exist a two black hole configuration in equilibrium. This problem has been open since the early days of General 
Relativity (see \citealp{Neugebauer:2013ee} for further references and
\citealp{Beig:1995fk, Beig:2008qi, Manko:2011qh} for 
different approaches and results on the subject). The Neugebauer and Hennig argument is the following: Start out with the spacetime metric
for an axially symmetric, stationary system containing two disconnected Killing horizons on the symmetry axis. Use the Ernst formulation 
\cite{Ernst68} to obtain a system of equations equivalent 
to  Einstein vacuum equations. Then the  inverse scattering method is used to build a unique and exact solution to the Ernst equations, known as
the double Kerr-NUT solution. A particular property of this solution is that both black holes can not satisfy the $A\geq8\pi|J|$ 
inequality simultaneously, which proves the non existence of two black 
holes in equilibrium. This result was generalized by \cite{Chrusciel:2011iv} to $I^+$ regular black 
hole spacetimes.

\section{Inequalities for objects}\label{sec:objects}

The inequalities presented in Sects.~\ref{sec:global} and \ref{sec:QL} are valid for black holes. The presence of 
such black hole is manifested 
through the hypothesis of the existence of a trapped surface or of a non-trivial topology in the initial data. 

The interest in geometrical inequalities for ordinary objects is twofold. The most basic question is whether Einstein equations set 
restrictions on the values that physical parameters for objects can attain. This is not the case in Newtonian theory unless some specific
matter model with intrinsic restrictions is used.  Is this the case in General Relativity? Are there some conditions on the mass, size, rotation, and charge of 
an object, such that if they are not fulfilled, the object can not exist within the theory? This is related to the second question we want
to address in this section. Are there geometrical inequalities for objects such that if violated, the object collapses to form a 
black hole? Clearly, the formation of a black hole after the collapse of an ordinary  object is one possible scenario leading to 
the non-existence raised in the first question. However, we emphasize that these two situations, i.e. \textit{an object exists and 
satisfies certain 
inequalities}, and \textit{an object does not exist because it forms a black hole} are in principle very different and require different treatments.

The problem of finding geometrical inequalities in the non-black hole setting is wide open. At this point it
 is not all that clear what kind of 
inequalities one should look for (some of them have been motivated in the introduction though, see Sect.~\ref{sec:newt}), nor what  
the proper systems and physical quantities are, that will produce  such inequalities (we discuss this point below). This makes 
research in this area look a bit erratic, where  new ideas are proposed or applied in almost every article. Because of this, 
we choose to present the results and discussions in a different manner as we do in previous sections.
We discuss the general problems first and then present the results with specific remarks for them.

\subsection{Discussion}\label{sec:discussionobj}

There are two major differences between non-black hole objects and black
holes. The first is the problem of how to characterize  the object in such a way that it produces  
the desired relation between physical 
quantities (like mass, electromagnetic charges or angular momentum) and size or shape parameters. That is, we would need some positivity 
condition to play the role of  the stability of MOTS or minimal surfaces used for obtaining black hole
inequalities (a non trivial topology for the underlying initial surface is not considered when studying physically reasonable ordinary
objects). 

The second problem, maybe less challenging but still open,
is how to properly measure the object.
When non-black holes objects are considered, one may want
to consider measures of 3-dimensional subsets of an initial data, and not just measures of 2-surfaces. This raises several difficulties as
there does not seem to exist consensus about what  the best or more appropriate measure is for the size of a non-black hole object. 
Indeed, a proper and suitable measure of size of an object should satisfy certain requirements. Namely, it should give a good, 
intuitive idea of size, it should be relatively easy to compute, it should be so chosen as to actually appear in the aimed geometrical inequalities.

These problems are aggravated by the fact that in general, there is not a special non trivial ordinary object known to saturate an estimate of the form
\be\label{generalobject}
[\mbox{Size}]\gtrsim \mbox{[Mass] or [Angular momentum] or [Charge]}
\ee 
where the symbol $[\cdot]$ indicates only the dependence of each term (by applying dimensional arguments one could propose a great number of more 
precise inequalities).
This leaves us
without a model solution to look at, as opposed to the extreme black holes in black hole inequalities. In fact, if such paradigmatic 
fully relativistic
object  satisfying certain geometrical inequality existed (as extreme Kerr--Newman black hole in the black hole scenario), it would give us
a path to what kind of inequality we should look for.

Note that to explore the rigidity case in \eqref{generalobject} means to address the problem of minimizing [Size] for given [Mass] 
(or given charge or angular momentum). Which in turn is closely related to the isoperimetric 
problem of minimizing area for given volume. 

As a measure of size in the left hand side of \eqref{generalobject}, one may attempt, inspired by the quasilocal inequalities for black 
holes (Sect.~\ref{sec:QL}), to use the area of the surface enclosing 
the object. However, there are counter examples to an inequality of the form $A\geq 4\pi Q^2$. The electrically counterpoised 
spheroids of dust, presented by \cite{bonnor98}, are regular, static, isolated systems that satisfy the energy conditions and 
whose enclosing surface can be made arbitrarily small relative to the charge enclosed, namely, $A<kQ^2$ for any positive, arbitrary 
number $k$.  Since these objects are highly prolate, it is 
expected that by assuming some kind of roundness on the enclosing surface, the area may give the desired estimate of charge. This example
does not mean that the area should not be considered as a measure of size for the not round enough objects. But it says that 
in some cases, the area alone is not enough to control the amount of charge the object can carry.

Taking this observation into account, two paths can be 
taken to arrive at  estimates of the form \eqref{generalobject}. One is to use special surfaces that are round enough. The other possibility is to 
use a measure that takes into account the deformation away from sphericity. In the first approach we encounter the following surfaces 
that capture the notion of round enough surface: isoperimetric surfaces 
(Sect.~\ref{sec:isop}) coordinate spheres and convex surfaces (Sect.~\ref{sec:ordinary}). Within the second approach, one may, as a 
first step, seek estimates using combinations of different well known measures, like area, distance to the boundary, etc. We come back to 
this point in Sect.~\ref{sec:ordinary}.

There is another important issue referred to ordinary objects that is closely related to black holes, and it is the question of the collapse of an
object to form a black hole. A few black hole formation criteria were constructed from geometrical inequalities stating that if certain
inequality is not satisfied, then a black hole is formed. We review them below.

\subsection{Results}\label{sec:ordinary}

As is the case for black holes, we divide the results according to whether angular momentum is considered explicitly into the inequality 
or not. This has to do mainly with the requirement of axial symmetry needed to define quasilocal angular momentum. As we see below, some of 
the results including angular momentum do not employ different methods to prove the inequalities.
Some results are quasilocal and some 
are global as they also incorporate the ADM mass. Different approaches have been taken to obtain the estimates.
However, no variational problem has been formulated. This, in particular, implies that there is no (non trivial) rigidity statement on the inequalities.

We mention here the setting where these inequalities are proven, and the main properties that lead to them.

\textit{Settings.} All inequalities presented in this section are proven for objects in an initial data set $(\Si, h_{ij}, K_{ij}, \mu,j^i)$.
The objects themselves are taken to be open, bounded regions $\Omega\subset\Si$ with smooth boundaries $\partial\Omega$. 

\textit{Inequality producer}. The various results we show in Sects.~\ref{objchar} and \ref{objang} use very different 
and apparently unrelated conditions that translate into the found inequalities. 
\begin{itemize}
 \item Strict positivity of the first eigenvalue of the linear differential operator $-\Delta+\frac{1}{2}{}^3R$.
\item Stability of minimal surfaces and MOTSs.
\item Stability of the quotient space of maximal slices in axial symmetry.
\item Positivity and monotonicity properties of the Geroch energy.
\item Stability of isoperimetric surface.
 \end{itemize}

\subsubsection{Inequalities for objects without angular momentum}\label{objchar}

\cite{schoen83d} study the black hole formation problem. They consider a maximal initial data 
$(\Si, h_{ij}, K_{ij}, \mu,
j^i)$ and an open subset $\Omega\subset\Sigma$ (the object) such that $\mu\geq \lambda>0$ on $\Omega$, where $\lambda$ is a constant. Then
\be\label{SY1}
\mathcal R_{SY}^2(\Omega)\leq\frac{\pi}{6\lambda},
\ee
where the radius $\mathcal R_{SY}(\Omega)$ is defined as follows. Take a simple closed curve $\Gamma$ in $\Omega$ which bounds a disk in $\Omega$. Let $r$ be the 
greatest distance from $\Gamma$ such that the set of all points within this distance form a torus embedded in $\Omega$. $\mathcal R_{SY}(\Omega)$ is 
the supremum of this $r$ over all curves $\Gamma$. 

The key point in this result is the fact that the first Dirichlet eigenvalue of the operator $-\Delta+\frac{1}{2}{}^3R$, if it is 
strictly positive, sets an upper bound to $\mathcal R_{SY}$.

Inequality \eqref{SY1} is purely quasilocal, the initial data does not need to be asymptotically flat. 

Shoen and Yau also obtain the following  black hole formation criterion: If $\Sigma$ is asymptotically flat and matter fields satisfy 
the energy condition $\mu-|j|\geq\lambda>0$ on $\Omega\subset\Si$, then the opposite inequality to \eqref{SY1} implies that $\Si$ 
contains an apparent horizon.

As stated by \cite{Murchadha86b}, the radius  $\mathcal R_{SY}$ captures the idea that the object must be large in every 
direction to avoid collapsing, but may be hard to compute in practical situations. He defines a new size measure \citep{Murchadha86b} 
as follows.  $\mathcal R_{OM}$ is the size of the largest stable minimal 2-surface $\Su$ that can be embedded in $\Omega$. By size 
we mean the   maximum of the distances (with respect to $h_{ij}$) from interior points  to the boundary 
$\Su$. The existence of such minimal surfaces is guaranteed when
$\Omega$ is mean convex (i.e., $\partial\Omega$ has positive mean curvature). He finds $\mathcal R_{OM}\geq \mathcal R_{SY}$ 
and obtains a sharpened version of \eqref{SY1}, that is
\be\label{OM1}
\mathcal R_{OM}^2(\Omega)\leq\frac{\pi}{6\lambda}.
\ee
\cite{Galloway:2008gc} generalize the above result to not necessarily maximal initial data and with MOTS replacing
the minimal surfaces. More precisely, they consider an object to be a relatively compact null mean convex open set $\Omega$  with connected 
boundary in an initial  data set $(\Si, h_{ij}, K_{ij}, \mu,
j^i)$. Define the radius of $\Omega$, $\mathcal R_{GOM}(\Omega)$ as the size of the greatest compact 
connected stable MOTS $\Su$ contained in $\Omega$ (size has the same meaning as in the O'Murchadha's definition of   $\mathcal R_{OM}$). Then, assuming $\mu-|j|\geq \lambda>0$ with $\lambda$ 
constant, obtain
\be\label{GOM1}
\mathcal R_{GOM}^2(\Omega)\leq\frac{\pi}{6\lambda}.
\ee
Note that the convexity condition is needed to guarantee the existence of the MOTSs in $\Omega$ \citep{Eichmair07}. 

We wish to remark that inequalities \eqref{OM1} and \eqref{GOM1} use stable minimal or trapped surfaces inside the object under study. 
In this way, they introduce the positivity condition which ultimately produces the inequality.

\cite{Reiris:2014tva} shows that the quotient space of maximal slices in axial symmetry satisfies a stability
property. This is an  interesting and strong argument which gives the 
desired positivity condition similar to that of stable minimal surfaces on the ambient space. The well known techniques of minimal 
surfaces are then adapted to obtain a similar bound
to that of Schoen and Yau in spherical symmetry. Namely, in spherically symmetric and asymptotically flat initial data
\be\label{R1}
\mathcal R_A^2(\Omega)\leq \frac{2\pi}{3\lambda}
\ee
where $\lambda$ is a positive constant bounding the energy density, $\lambda\leq\mu$, and $\mathcal R_A(\Omega)$ is the areal radius of the constant radius sphere $\Omega$. 

This result gives also the following black hole formation criterion: If the 
energy density of the object satisfies $\rho>\pi/6M^2$ where $M$ is the ADM mass, then the object lies inside a black hole and is not in
static equilibrium.

He also obtains
\be\label{R3}
\mathcal R_A\geq \frac{Q^2}{2M_{ADM}}
\ee
for spherically symmetric, asymptotically flat initial data, satisfying the dominant energy condition.

\cite{Khuri:2015xpa} decomposes the matter density as an electromagnetic part (subindex $EM$) and a non-electromagnetic part (subindex $M$) as 
$\rho=\rho_{M}+\rho_{EM}$ and $j=j_{M}+j_{EM}$ and assumes that the non-electromagnetic part satisfies the dominant energy condition, that is
$\mu_M\geq|j_M|$. Moreover, $\mu_M$ is taken to be constant. Then from this condition and the definition of electromagnetic charge he obtains 
\be\label{K2}
Q^2\leq\frac{A}{2\pi}\int_{\partial\Omega}\mu_M-|j_M|
\ee
where $A$ is the area of $\partial\Omega$. Then, using Shoen and Yau bound \eqref{SY1} he obtains
\be
|Q|\leq\frac{A}{\sqrt{12}\mathcal R_{SY}}
\ee
Moreover, if $\Omega$ is mean convex, then the same bound holds for $\mathcal R_{OM}$ instead of $\mathcal R_{SY}$.

\cite{Anglada:2015tan}  study the spherically symmetric, electrovacuum case. They find that if the initial data 
is asymptotically flat and spherically symmetric, and 
if outside a ball $\Omega$ with finite areal radius $\mathcal R_A$, it is electrovacuum and untrapped, then
\be\label{DQ}
\mathcal R_A\geq\frac{|Q|}{2}
\ee
This inequality is weaker than \eqref{R3} and it is saturated at $Q=0$ with vanishing radius and total mass. This inequality is not quasilocal in the sense that asymptotic 
flatness is required for \eqref{DQ} to hold and, in fact there are non asymptotically flat examples were it is violated. Note that \eqref{DQ} does not
use the bound \eqref{SY1} nor the radius $\mathcal R_{SY}$, as $\mathcal R_A$ is a more natural size measure in spherical symmetry (see 
the discussion about the convenience of using the surface area as a size measure for objects, in Sect.~\ref{sec:discussionobj}).

\subsubsection{Inequalities for objects with angular momentum}\label{objang}

The key ingredient needed to include the angular momentum into a geometrical inequality is to relate it with the Einstein constraints. 
This is done via the relation with the current density $j^i$ or with the extrinsic curvature $K_{ij}$ of $\Si$.

\cite{Dain:2013gma} considers maximal axially symmetric initial data
$(\Si, h_{ij}, K_{ij}, \mu,
j^i)$ with constant energy density $\mu$ and non vanishing current density $j^i\neq0$, satisfying the dominant energy 
condition. He takes an  object to be an axially symmetric  open subset $\Omega$ of $\Sigma$.  From the definition  of angular 
momentum (see equation \eqref{eq:70}) he bounds the angular momentum $J$ of $\Omega$ in terms of the integral of the current 
density (and via the energy condition, in terms of $\mu$) and the norm of the Killing  vector $\eta^i$ associated to the axial symmetry, that is $\eta=\eta^i\eta_i$,
\be
|J|\leq\int_\Omega|j|\sqrt{\eta}\leq\mu\int_\Omega\sqrt{\eta}.
\ee
Then, using the Hamiltonian constraint he obtains $R\geq16\pi\mu$, and therefore, \eqref{SY1} gives the geometrical inequality between the 
angular momentum $J$ of $\Omega$ and the Shoen and Yau size.
\be\label{D1}
|J|\leq \frac{\pi}{6}\frac{\int_\Omega\sqrt{\eta}}{\mathcal R_{SY}^2}.
\ee
In fact, Dain proposes the right hand side of \eqref{D1} as a new measure of size.

By assuming $\Omega$ to be mean convex, the same inequality is obtained when  $\mathcal R_{OM}$ is used in the definition of $\mathcal R_D$

\cite{Khuri:2015zla} extends this result to not necessarily maximal initial data satisfying a stronger version of the dominant energy
condition, namely $\rho\geq |\bar j| +|j_\eta|$ where $j_\eta$ is the current density in the direction of the axial Killing 
vector field $\eta^i$ and $\bar j$ is the current in the orthogonal directions. More precisely, Khuri considers an axially symmetric initial data,  without compact apparent horizons, which is 
asymptotically flat or has a strongly untrapped boundary, then for an open set $\Omega$ in the initial data it holds
\be\label{K1}
|J|\leq \frac{3\pi C_0}{16}\frac{\int_\Omega\sqrt{\eta}}{\mathcal R_{SY}^2}.
\ee
where $C_0:=\max_\Omega(\mu-|\bar j|)/\min_\Omega(\mu-|\bar j|)$.

This result gives the following black hole formation criteria: Given an axially symmetric initial data such that
the initial surface $\Sigma$ is asymptotically flat or has a strongly untrapped boundary. Under the energy condition stated above, if there exists a bounded
region $\Omega$ where \eqref{K1} does not hold, then $\Sigma$ contains an apparent horizon.

Without using the Schoen and Yau bound \eqref{SY1}, \cite{Reiris:2014tva} goes a different route to obtain geometrical inequalities for objects. Using techniques of minimal surfaces he finds estimates on the shape of an axially symmetric object in terms of the angular momentum when the object does not 
intersect the symmetry axis and is connected:
\be\label{R2}
|J|\leq\left(1+\frac{P}{\pi D}\right)\frac{\pi}{2}\mathcal R_C^2
\ee
where $P$ is the transversal perimeter of $\Omega$, $D$ is the distance from $\Omega$ to the symmetry axis and $2\pi\mathcal R_C$ is the length 
of the greatest axisymmetric orbit in $\Omega$. Note that this result gives information, not only  about the size, but about the shape of the 
object. This implies that in order to control the angular momentum of an ordinary object, size in all directions should be 
considered, an observation that may also be valid for electrically charged ordinary objects (see the discussion about the 
Bonnor example mentioned in Sect.~\ref{sec:discussionobj}).

The inequalities \eqref{D1},  \eqref{K1} and  \eqref{R2} suggest the existence of an appropriate size measure, $\mathcal R$, probably 
defined  in terms of the norm of the Killing vector $\eta$, as well as measures in relevant spatial directions, such that
an inequality of the form
\be
|J|\lesssim\mathcal R^2
\ee
holds for ordinary objects.

We finally present an inequality that is global in the sense that in includes the ADM mass.

Using the inverse mean curvature flow on asymptotically flat, axially symmetric initial data $(\Si, h_{ij}, K_{ij}, \mu,
j^i)$, \cite{Anglada:2016dbu} study convex regions $\Omega$ where the current density has compact support. Assuming that the initial data 
satisfies the dominant energy condition and has no minimal surfaces, they find 
\be\label{G1}
M_{ADM}\geq m_T+\frac{J^2}{5\mathcal R_A\mathcal R_c^2}
\ee
where $\mathcal R_A$ and $\mathcal R_c$ are the areal and circumferential radius of the convex flow surface $S_T$ such that $S_t$ is convex for 
$t\geq T$. Also, $m_T$ is a positive constant
\be
m_T:=\frac{1}{16\pi}\int_0^{\mathcal R_A}d\xi\int_{S_\xi} RdS
\ee
and $\xi$ is the areal radius coordinate and $R$ is the curvature scalar of $h$. 

The positivity and monotonicity properties of the Geroch energy are crucially used to relate the ADM with the curvature scalar and the 
norm of the Killing vector field, $\eta$, on the surfaces
defined by the flow. Using the Hamiltonian constraint together with the definition \eqref{eq:38}, the
 scalar curvature is bounded by the angular momentum of the surfaces. Finally convexity of the flow surfaces is used to control the 
 evolution of  $\eta$ along the flow.

Inequality \eqref{G1} also gives information about the shape of the object. It says that if the total mass is fixed, 
then the angular momentum determines how oblate or prolate the object can be. We also notice that the term $m_T$ plays the role of a 
quasilocal mass (see \citealp{Malec:2002ki}). 
 
It is remarkable that the inequalities obtained in this section, although with different technical conditions, give rise to inequalities
similar to the ones discussed in Sect.~\ref{sec:newt}, which were informally derived from Newtonian considerations and the condition that
nothing travels faster than light.

\subsubsection{Isoperimetric surfaces}\label{sec:isop}

As we mention in Sect.~\ref{sec:BH}, isoperimetric surfaces in initial data are  an important system from which relations between 
physical and geometrical quantities can be obtained. We refer the reader to the articles by \cite{Eichmair:2012tz, Eichmair:2011sx} for a detailed account on the results related to isoperimetric surfaces in Riemannian manifolds with application 
to General Relativity. See also Sect.~\ref{sec:BH} for references on discussions about the Penrose inequality for isoperimetric surfaces.
           
In this section we focus on inequalities relating size, angular momentum and charges, so, in this sense, they are quasilocal inequalities.

\cite{Dain:2011kb} study electro-vacuum, maximal initial data, possibly with a non negative cosmological 
constant and find that if $\Su$ is a  stable isoperimetric
sphere, then
\begin{equation}
  \label{eq:isoperimetric} 
A(\Su)\geq \frac{4\pi}{3} Q^{2}(\Su). 
\end{equation}  
Stability here means that the area function is not only critical  at the isoperimetric surface $\Su$, but also a minimum.

\cite{Acena:2012wg}, characterize the
behavior of isoperimetric surfaces in Reissner--Nordstr\"om and find, among other results that the spheres $r=constant$ in the 
Reissner--Nordstr\"om metric  are isoperimetric
  stable for $0\leq|Q|\leq M$ and satisfy the bound
\begin{equation}
  \label{eq:2x}
 A \geq \frac{16}{9}\pi Q^2.  
\end{equation} 
Moreover there is not a sphere in Reissner--Nordstr\"om where the inequality \eqref{eq:2x}  is
  saturated. The inequality is saturated in the limit   when the extreme case is approached from the superextreme case.

Up to now, the only result involving angular momentum for isoperimetric
surfaces is proven by \cite{Reiris:2014tva}.  Let $\Su$ be a stable isoperimetric, axisymmetric sphere enclosing an object 
$\Omega$ (and
nothing else). Then,
\begin{equation}
  \label{eq:2}
  |J|\leq c_1 R \sqrt{A} \leq c_22 R L
\end{equation}
where $c_1 = 6/(8\pi3/2)$, $c_2 = 6/(4\pi)$, $∣J∣$ is the angular momentum of $\Omega$ and $A$, $R$ and $L$ are,
respectively, the area of $\Su$, the length of the greatest axisymmetric orbit in $\Su$
and the distance from the North to the 
South pole of $\Su$.


\section{Open problems}\label{sec:open}
There are a number of open problems that need to be addressed in order to understand more completely the type of estimates one can
obtain both for black holes and ordinary objects. Most of them were mentioned and/or  discussed in the appropriate section. We list them 
below as well.

\begin{itemize}

\item \textit{Removing axial symmetry}. Axial symmetry is a requirement in all the geometrical inequalities involving angular momentum presented 
in this article. However, deviations from axial symmetry are of major importance, especially in astrophysics and numerical simulations. 
Moreover, there seems to not be any physical reason why the inequalities presented in this article should not hold in more general, non
axially symmetric systems. In the case of quasilocal inequalities outside axial symmetry, it is worth recalling that the variational 
problem for the mass functional $\mathbb M$ presented in Sect.~\ref{sec:proofAJ} also holds for non axially symmetric functions. This 
shows the major role that extreme Kerr--Newman black holes plays as a limit solution, among a wider class of not necessarily axially symmetric
solutions.

\item \textit{Mass-Angular momentum inequality for data with inner boundary}. We have seen in Sect.~\ref{sec:MQ} that the Mass-Charge 
inequality can be formulated in terms of initial data where  the initial surface $\Si$ is either complete with non trivial topology, 
or has a weakly trapped inner boundary. On the other hand, the Mass-Angular momentum inequalities presented in Sect.~\ref{subsec:MJ} are 
proven only for complete initial surfaces with non trivial topology. Extending this result to manifolds with boundaries is important 
for three reasons: First it would complete and unify the results about this type of global inequalities. Second, the proper formulation and 
resolution of the variational problem needed to derive the desired inequality when an inner boundary is present (analogous to the one used
in the proof for the case of complete manifold), would  clarify the role that extreme black holes play as borderline solutions. Finally, it
seems that if one wants to make a connection between this geometrical inequality and the Penrose inequality, (see next item) a careful understanding of this
case may be of use.

\item \textit{Connection with Penrose inequality}. The connection of the geometrical inequalities presented in this article and the 
positive mass theorem and the Penrose inequality seems to become deeper as further studies are performed. Not only they involve the 
 same physical and geometrical quantities, i.e., Mass, Area (or Size), Angular momentum, Charge, but also the techniques used in both 
problems seem to not be so different (see Sect.~\ref{sec:ordinary} for an inequality for objects using the inverse 
mean curvature flow). Exploring this connection may shed light into the problems and possible resolutions. In this respect, see the article
by \cite{Anglada:2017ryp}, where he adapts the results in \cite{Anglada:2016dbu} for ordinary objects to study the Penrose inequality.

\item \textit{Minimum of mass functional for multiple rotating black holes}. The global inequality for multiple black holes, 
\eqref{eq:49a} is 
written in terms of the value of the mass functional on a minimizer solution. This minimizer is not known explicitly and moreover, it is 
not even known the expected outcome, as some numerical simulations suggest different results. Obtaining an explicit 
form of this value is of great importance because it would tell us whether the total mass controls the individual angular momenta of 
the black holes, or the total angular momentum of the system, and exactly how it does it.

\item \textit{Global inequality with $\Lambda$}. Global inequalities relating mass, angular momentum and/or charge that also include explicitly  a 
cosmological constant $\Lambda$ have not been proven yet. A negative cosmological constant, however, has been admitted in the statements of the
Mass-Angular momentum-Charge 
 inequalities \citep{Cha:2017gej}.

\item \textit{Connection between $\mM$ and $\mathbb M$.} Another issue that must be better understood is the connection between the 
global and quasilocal inequalities for black holes that include the angular momentum. A partial implication was presented in section 
\ref{relation} but there are many issues that are not entirely clear yet. This is not an easy problem
since it involves relating global and quasilocal settings. Its full resolution might give a hint into the connection  
with the Penrose inequality.

\item \textit{Quasilocal estimate with negative $\Lambda$}. The way the cosmological constant appear into the mass functional 
$\mathbb M$ makes the procedure used to prove the Area-Angular momentum-Cosmological constant inequality hard to adapt when $\Lambda$ is 
negative. Note that the problem with the negative $\Lambda$ is not 
about how it enters into a generalization of the Area-Angular momentum inequality, but about whether such an inequality does exist. The 
works mentioned in Sect.~\ref{sec:Lambda} suggest that it does exist and a particular inequality motivated by extreme Kerr--Newman AdS
black hole has been proposed. This problem is far from solved and new techniques must be implemented.

\item \textit{Ordinary objects}. As was discussed in Sect.~\ref{sec:objects}, there are very basic questions that are unanswered 
with respect to geometrical inequalities for objects. Things like: what inequality  we expect to obtain, how we should characterize 
the object and how we should measure them, are not clear. Concerning the first two issues, it is crucial to understand in what 
class of  ordinary objects one expects to obtain a  geometrical 
inequality. By this we mean that a positivity condition seems to be needed in order to derive the desired estimate. 
This leads naturally to 
the following question: Do all objects, say, in axial symmetry for simplicity, have a restriction on the allowed valued of their parameters?
In particular, should they be \textit{round enough}? 

\item \textit{Measure of size for ordinary objects}. This issue was discussed in Sect.~\ref{sec:objects}. There are various alternative 
notions of size but more work needs to be done. As seen in the results presented in Sect.~\ref{objang}, in the case of axial symmetry, 
it may be convenient to study measures constructed from the norm of the axial Killing vector field, $\sqrt{\eta}$. This is supported by the following 
observations: The norm $\sqrt{\eta}$ is bounded by the equatorial radius 
$\mathcal R_c$ (defined as the length, divided by $2\pi$ of the greatest axially symmetric circle. This gives a clear and natural 
measure of size relevant for rotating objects. Also, for convex surfaces the variation of $\eta$ 
along the inverse mean curvature flow is controlled by $\eta$ itself. Also, the measure should take into account deviations 
from sphericity in all directions.

\item \textit{Connection with Hoop conjecture.} Some versions of the Hoop conjecture suggest to look for geometrical inequalities relating 
size, angular momentum and some measure of quasilocal mass of a certain region of spacetime \citep{Senovilla:2007dw}. There are several
different quasilocal masses in the literature \citep{Szabados04}, but the problem of identifying the appropriate one(s) that simultaneously capture 
the matter content of the region, and that give rise to the desired meaningful geometrical inequalities is still open.

\end{itemize}

\begin{acknowledgements}
We wish to thank L\'{a}szl\'{o} Szabados for fruitful discussions and Jos\'e Luis Jaramillo for careful comments about some aspects of the manuscript. 
This work was supported by Consejo Nacional de Investigaciones Cien\'ificas y Tecnol\'ogicas and Secretar\'ia de Ciencia y T\'ecnica-
Universidad Nacional de C\'ordoba.
\end{acknowledgements}


\bibliographystyle{spbasic}      
\bibliography{biblio}   

\end{document}